\documentclass[aps,prl,twocolumn,superscriptaddress,amssymb]{revtex4-2}
\usepackage[a4paper, total={7in, 10in}]{geometry}
\usepackage{graphicx} 
\graphicspath{{Figures/}}
\usepackage{amsmath}
\usepackage{physics}
\usepackage{dsfont}
\usepackage{dcolumn}
\usepackage{xcolor}
\usepackage{bbold}
\usepackage[normalem]{ulem}

\usepackage{appendix}
\usepackage{units}
\usepackage[caption=false]{subfig}
\usepackage{lipsum}
\usepackage{graphicx}
\usepackage{pgfplots}
\pgfplotsset{compat=1.18} 
\usepackage[hidelinks]{hyperref}
\usepackage{cleveref}
\usepackage[T1]{fontenc}
\hypersetup{colorlinks=true,linkcolor=blue}

\newcommand{\beginsupplement}{
  \clearpage
  \onecolumngrid
  \setcounter{page}{1}
  \setcounter{section}{0}
  \setcounter{figure}{0}
  \setcounter{table}{0}
  \setcounter{equation}{0}
  \renewcommand{\thepage}{S\arabic{page}}
  \renewcommand{\thesection}{S\arabic{section}}
  \renewcommand{\thefigure}{S\arabic{figure}}
  \renewcommand{\thetable}{S\arabic{table}}
  \renewcommand{\theequation}{S\arabic{equation}} % <-- equations now S#
}

\begin{document}

\title{
Probing Electrostatic Disorder via g-Tensor Geometry
}

\author{Edmondo Valvo}
\email{ evalvo@tudelft.nl}
\affiliation{QuTech and Kavli Institute of Nanoscience, Delft University of Technology, PO Box 5046, 2600 GA Delft, The Netherlands
}

\author{Christian Ventura-Meinersen}
\affiliation{QuTech and Kavli Institute of Nanoscience, Delft University of Technology, PO Box 5046, 2600 GA Delft, The Netherlands
}

\author{Michèle Jakob}
\affiliation{QuTech and Kavli Institute of Nanoscience, Delft University of Technology, PO Box 5046, 2600 GA Delft, The Netherlands
}

\author{Stefano Bosco}
\affiliation{QuTech and Kavli Institute of Nanoscience, Delft University of Technology, PO Box 5046, 2600 GA Delft, The Netherlands
}
\author{Tereza Vakhtel}
\affiliation{QuTech and Kavli Institute of Nanoscience, Delft University of Technology, PO Box 5046, 2600 GA Delft, The Netherlands
}
\author{Maximilian Rimbach-Russ}
\affiliation{QuTech and Kavli Institute of Nanoscience, Delft University of Technology, PO Box 5046, 2600 GA Delft, The Netherlands
}

\date{\today}

\begin{abstract}
 Low-frequency charge noise induced by fluctuating electrostatic disorder is a major limitation for semiconductor hole spin qubits. Here, we analyze the quasistatic response of a hole spin qubit to individual two-level fluctuators (TLFs). We show that, due to the anisotropy of the g-tensor, the qubit response depends on the geometry of the fluctuator-induced dipolar perturbation. We then propose a readout protocol that isolates selected g-tensor components through an accumulated Berry phase and estimate, within our readout model, an order-unity signal-to-noise ratio with a total protocol time in the tens of microseconds. Finally, using microscopic simulations, we compute the quantum Fisher information (QFI) to identify magnetic field directions and confinement regimes in which the qubit is most sensitive to disorder-induced variations of selected g-tensor components.

\end{abstract}

\maketitle

%\maketitle

\textit{Introduction --- }
Germanium quantum dot-based hole-spin qubits in strained Ge/SiGe heterostructures are emerging as a promising candidate for condensed-matter based quantum computing because of the material’s small effective mass granting large orbital splitting, compatibility with standard silicon nano-fabrication techniques ~\cite{lossQuantumComputationQuantum1998,burkardSemiconductorSpinQubits2023,stanoReviewPerformanceMetrics2022,vandersypenInterfacingSpinQubits2017,philipsUniversalControlSixqubit2022,xueQuantumLogicSpin2022,millsTwoqubitSiliconQuantum2022,zwerverShuttlingElectronSpin2023,desmetHighfidelitySinglespinShuttling2025,takedaQuantumErrorCorrection2022,noiriShuttlingbasedTwoqubitLogic2022,wuSimultaneousHighFidelitySingleQubit2025,steinackerIndustrycompatibleSiliconSpinqubit2025,gonzalez-zalbaScalingSiliconbasedQuantum2021,petitUniversalQuantumLogic2020,siegelQuantumSnakesPlane2026,PhysRevA.109.032433,svastitsReadoutSweetSpots2026,liFlexible300Mm2020,kloeffelProspectsSpinBasedQuantum2013,fangRecentAdvancesHolespin2023,maurandCMOSSiliconSpin2016,tidjaniThreeDimensionalArrayQuantum2025} and a  strong spin-orbit interaction that allows for all-electrical control, with single and two-qubit fidelities above 99\%~\cite{zhangUniversalControlFour2025,vanriggelenPhaseFlipCode2022,Wang2024Hopping,hendrickxFourqubitGermaniumQuantum2021,jirovecSinglettripletHoleSpin2021,wangUltrafastCoherentControl2022,liuUltrafastElectricallyTunable2023,lawrieSimultaneousSinglequbitDriving2023,Scappucci2020,Stehouwer2025,vorreiterPrecisionHighspeedQuantum2025,jirovecManybodyInterferometrySemiconductor2026,saez-mollejoExchangeAnisotropiesMicrowavedriven2025,johnTwodimensional10qubitArray2025a,jirovecMitigationExchangeCrosstalk2025,ivlevOperatingSemiconductorQubits2025,farinaSiteresolvedMagnonTriplon2025}. Moreover, extensive work has characterized the full g-tensor\cite{Terrazos2021,adelsbergerHolespinQubitsGe2022,seidlerSpatialUniformityGtensor2025,Wang2024Modelling} and exploited its anisotropy and electric tunability for optimal operation \cite{wangOptimalOperationPoints2021,Bosco2021,bassiOptimalOperationHole2026,rimbach-russGaplessSingleSpinQubit2025,boscoExchangeOnlySpinOrbitQubits2024,PRXQuantum.2.010348}and
coherence sweet-spots \cite{Hendrickx2024, mauroGeometryDephasingSweet2024,wangDephasingPlanarGe2025}.

However, the same electric susceptibility that gives rise to fast manipulation also couples the spin to stray charges. This presents a major challenge for wafer scale integration because it creates device‑to‑device variability of $g$‑factors, Rabi frequencies and coherence sweet spots~\cite{Martinez2022,martinezVariabilityHolespinQubits2026,variability} along with significant complications for optimal control protocols~\cite{glaserTrainingSchrodingerCat2015,theisCounteractingSystemsDiabaticities2018,kochQuantumOptimalControl2022,barnesDynamicallyCorrectedGates2022,rimbach-russSimpleFrameworkSystematic2023,ventura-meinersenQuantumGeometricProtocols2025,meinersenUnifyingAdiabaticStatetransfer2025,ventura-meinersenMultilevelSpectralNavigation2026} that can be attributed to sub-nm differences in the local electrostatic environment~\cite{stehouwerGermaniumWafersStrained2023,sangwanImpactSurfaceTreatments2025,paqueletwuetzReducingChargeNoise2023,elsayedLowChargeNoise2024,kellyIdentifyingMitigatingErrors2025,Wang2024Hopping,vanriggelen-doelmanCoherentSpinQubit2024,seidlerSpatialUniformityGtensor2025,tosatoCrossbarChipBenchmarking2026}. 
Recent works in platforms with similar heterostructures ~\cite{Choi2024,berrittaPhysicsinformedTrackingQubit2024,Park2025,benestadAutomatedSituOptimization2025,berrittaEfficientQubitCalibration2025} have demonstrated active and passive manipulation of TLFs, however, these techniques require fast and directional TLF sensing~\cite{patomakiElongatedQuantumDot2024,kanaar2025fastchargenoisesensing}. 
\begin{figure}[h!]
    \centering
    \includegraphics[width=\linewidth]{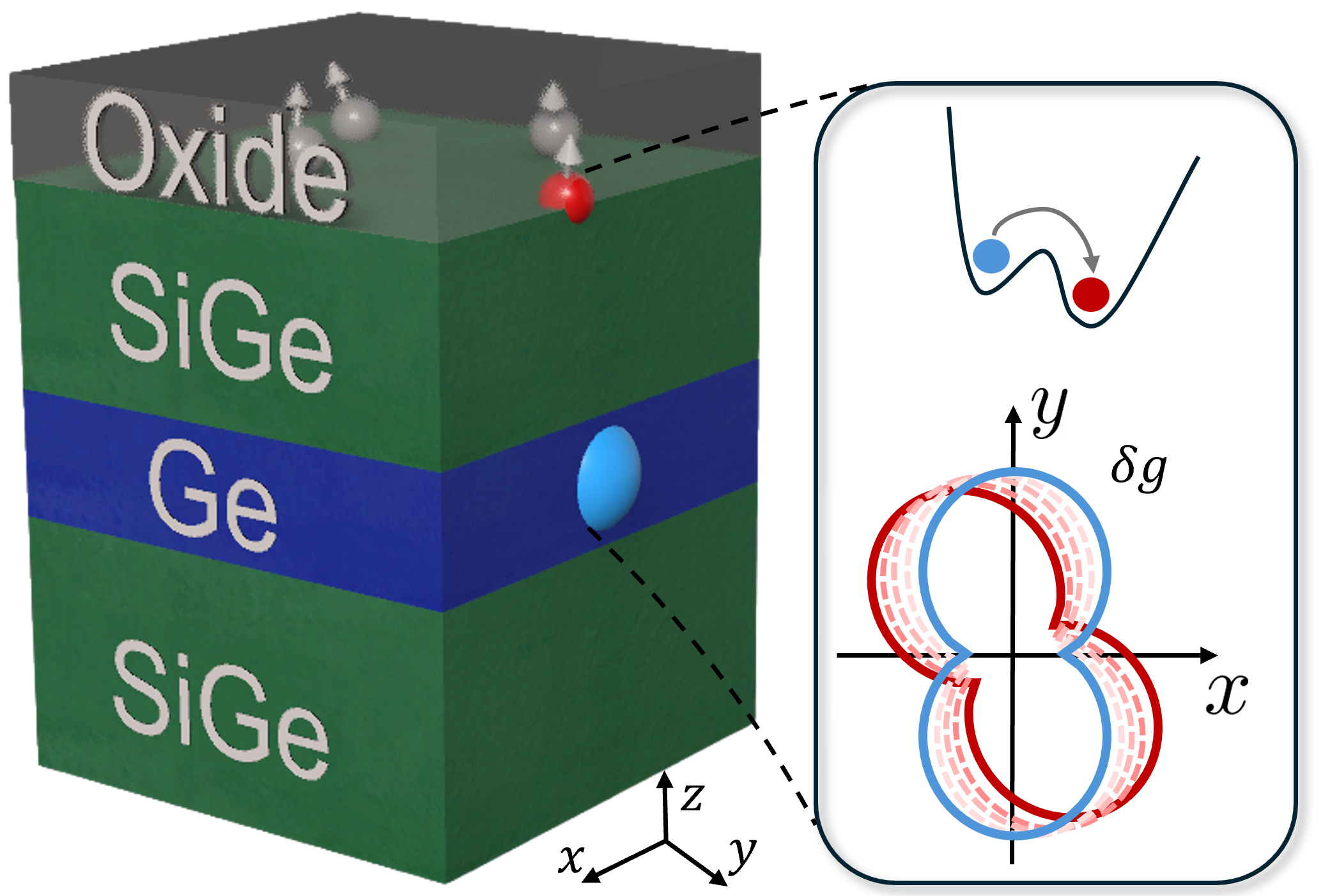}
    \captionsetup{justification=raggedright,singlelinecheck=false}
    \caption{Highlighted in red is the active TLF, modeled as an effective dipolar electrostatic perturbation that modifies the quantum-dot environment. The fluctuating trap is placed at the SiGe/Oxide interface and is surrounded by other remote fluctuators. In the top inset, switching between the reference and metastable configurations of the active fluctuator is illustrated. 
    The quantity $\delta g$ is defined as the variation to the g-tensor resulting from the electrostatic deformation of the TLF. }
    \label{fig:Illustrative-Figure}
\end{figure}
\begin{figure}[htbp]
  \centering
  \includegraphics[width=\linewidth]{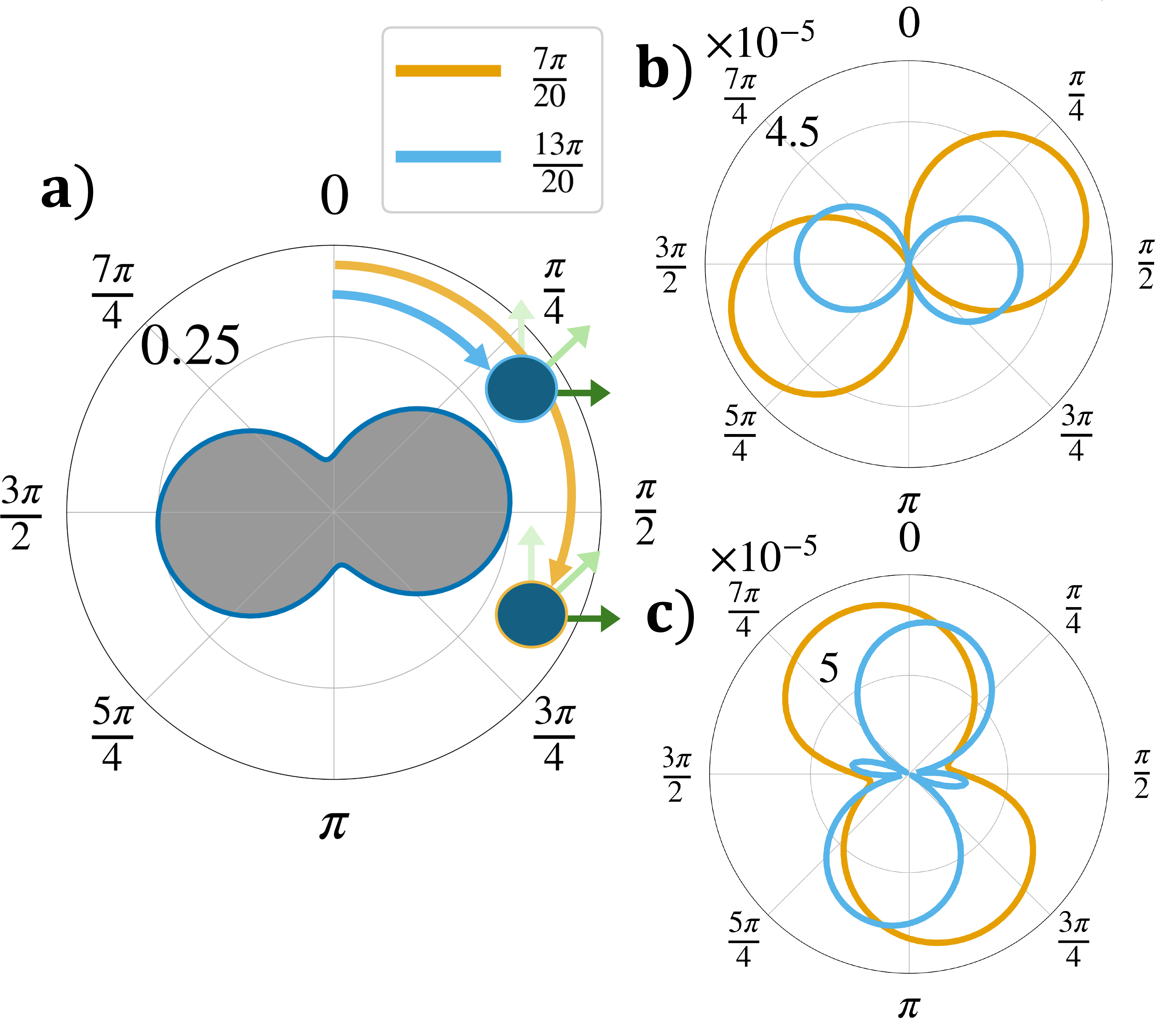}
      \captionsetup{justification=raggedright,singlelinecheck=false}
  \caption{a) The trap is placed in the top-right quadrant, at $63^\circ$ (blue arrow) from the x-axis for configuration 1. The green color-graded arrows indicate the possible displacement direction of the TLF $\delta \textbf{r}$, generating a unique dipolar field. The second configuration is for a TLF placed at $117^\circ$ (orange arrow) from the x-axis.  b) Variation of the g-factor $\delta g$ for a tilting angle $\phi_\text{Tilt}=0$ for the two trap placements and displacement direction aligned along the x-axis . c) Variation of the g-factor for a tilting angle $\phi_\text{Tilt}=\pi/2$ for the two trap placements and displacement direction aligned along the x-axis. The g-tensor variations follow the confinement rotation but presents a lack of periodicity given by a local strain fluctuation. In c) the lobe of the $\frac{13 \pi}{20}$ scenario is due to the $\delta g$ acquiring negative values.}
  \label{fig:sixpanel}
\end{figure}
To this end, hole qubits offer an intrinsic advantage because the g-tensor is both highly anisotropic and depends on the electrostatic environment~\cite{Scappucci2020,fangRecentAdvancesHolespin2023}. The presence of a TLF will result in a specific quasistatic configuration of the electrostatic landscape that translates into a small g-tensor modulation dependent on the fluctuator's magnitude and spatial orientation.\par

In this work, we highlight this susceptibility and extract the microscopic response of the Ge hole qubit to the quasistatic potential generated by a TLF. We note that our study can be generalized to hole spin qubits in Silicon\cite{lilesElectricalControl$g$2021,lilesSinglettripletHolespinQubit2024,jinProbingGtensorReproducibility2024}, different charge models, and positions. To extract the microscopic effect of these fluctuators, we present a geometric readout scheme that converts the modification of individual components of the g-tensor, due to a quasistatic configuration of the TLF, into a measurable Berry phase, while strongly suppressing dynamical contributions with dynamical decoupling. In particular, whereas a simpler frequency-only measurement probes the total Zeeman splitting and therefore mixes the response of different $g$-tensor components, the present protocol is directly sensitive, to linear order, to a small $g$-tensor component that couples an in-plane magnetic field to an out-of-plane effective field, thereby providing information complementary to frequency-only readout. Moreover, it does not require the usage of a vector magnet. We estimate that a signal-to-noise ratio of order one can be reached within integration times of hundreds of nanoseconds, and the total protocol duration remains in the tens of microsecond range. To quantify the sensitivity of the hole qubit to parameter variation caused by charge fluctuations, we employ the QFI ~\cite{forghieriQuantumEstimationRemote2023}.

\textit{Microscopic Model --- }
We model the germanium hole qubit with the Hamiltonian $H_0=H_{\text{LK}}+H_{\text{BP}}+V(\textbf{r})$, where $H_{\text{LK}}$ is the 4-band Luttinger-Kohn Hamiltonian and $H_{\text{BP}}$ is the 4-band Bir-Pikus Hamiltonian comprising both biaxial and shear strain components (for further details see SM \cite{SM}). The last term $V(\textbf{r})$ describes the electrostatic confinement, which we assume to be approximately separable, $V(\textbf{r})=V_{xy}+V_z$ . In particular, we choose
$
V_{xy}(x,y) =  \left(\frac{m \omega_x^2}{2} x^2 + \frac{m \omega_y^2}{2} y^2\right)
$
to be a harmonic in-plane confinement with effective mass $m=m_0/\sqrt{\gamma_1^2 - \gamma_2^2}$, where $\gamma_1$ and $\gamma_2$ are Luttinger parameters. The additional linear potential due to the plunger-gate electric field, together with the step-like heterostructure confinement, is included through piecewise-defined potential terms $V_z$, see the  Supplemental Material (SM)~\cite{SM}. We project this Hamiltonian onto an analytical basis set composed of in-plane harmonic eigenstates and a Gaussian approximation of the Airy wavefunctions in the growth direction~\cite{Wang2024Modelling,wangDephasingPlanarGe2025,variability}. Finally, we introduce the disorder as distinct charge defects and model them as a displaced Coulomb potential that modifies the electrostatic confinement as $V(\textbf{r})\xrightarrow[]{}V(\textbf{r})+V_c(\textbf{r)}$ \footnote{We model the charges as dipoles in one metastable state and consider the other state as the complete absence of the charge. However, a full electrostatic landscape of screened monopoles is added, and the modelling approach is explained in further detail in the SM ~\cite{SM}. The specific effects of these monopoles are included in the term $g_C$ of the g-tensor corrections \cref{g tensor corrections}.} with
\begin{equation}\label{Coulomb}
    V_c(\textbf{r})=\sum_j \frac{F_c}{|\textbf{r}_j+\delta\textbf{r}_j|} - \frac{F_c}{|\textbf{r}_j|} - \frac{F_c}{|\textbf{r}_j+\delta\textbf{r}_j + \textbf{r}_m|} + \frac{F_c}{|\textbf{r}_j+\textbf{r}_{m}|} ,
\end{equation}
with prefactor $F_c=\frac{e}{4 \pi \epsilon_0 \epsilon_r}$, where $\epsilon_0$ and $\epsilon_r$ are the vacuum and relative germanium permittivity respectively. Here $\textbf{r}$ is the position of the defect, $\delta\textbf{r}$ describes the displacement of the fluctuator and $\textbf{r}_m$ is the position of the mirror charge modelling the screening of the top gate at $7$nm from the SiGe/Oxide interface. Accordingly, the active TLF is modeled as an effective dipolar perturbation on top of a reference electrostatic landscape, rather than a full charge switching between a bistable potential.\newline
\par 
Subsequently, we define a magnetic perturbation $H_B= H_{\text{Zeeman}} + H_o $, where the first term is the anisotropic Zeeman Hamiltonian $H_{\text{Zeeman}}=2 \mu_B \textbf{B}\cdot(\kappa \textbf{J}  + q \textbf{J}^3)$ representing the interaction between the hole qubit and the magnetic field. The second term, that we shall denote as $H_o=- 2\mu_B [\gamma_3\{\textbf{A}\cdot\textbf{J},\textbf{k}\cdot \textbf{J}\}+(\gamma_2-\gamma_3)\{\textbf{A},\textbf{k}\}\cdot\textbf{J}^2]$, contains the linear terms in magnetic field coming from the orbital contributions of the Luttinger-Kohn Hamiltonian after the dynamical momentum substitution $\hbar\textbf{k}\to\pmb{\pi}=\hbar\textbf{k} +e \textbf{A}$. To avoid numerical instability due to the finite basis set dimension, we choose a gauge that excludes explicit z dependence $A=(- \frac{B_z y}{2},\frac{B_z x}{2},-B_y x+ B_x y)$. The linear response to an applied magnetic field is then extracted according to the \textit{g-tensor} formalism ~\cite{Venitucci2018} by projecting the magnetic field-dependent perturbation $H_B$ on the unperturbed eigenstates, obtained via numerical diagonalization, yielding a $3 \times 3$ matrix known as the g-tensor. \par Considering the various terms in the Hamiltonian $H_0$, the $g$-tensor can be decomposed into the following contributions
\begin{equation} \label{g tensor corrections}
    g=\underbrace{g_Z+g_\varepsilon+g_C}_{g_0} + \delta g,
\end{equation}

where $g_Z$ is the contribution coming from the bare Zeeman Hamiltonian, while $g_\varepsilon+g_C$ are renormalizations coming from the strain tensor and the confinement potential in the three spatial directions, respectively \cite{abadillo-urielHoleSpinDrivingStrainInduced2023,rimbach-russGaplessSingleSpinQubit2025,variability}. Finally, $\delta g$ denotes the correction to the reference $g$-tensor induced by the active TLF. 

\par \textit{Charge Sensing --- }
To engineer a setting for a preferential directional sensitivity to the TLFs, we squeeze and tilt the planar confinement \cite{Bosco2021}. Practically this is achieved by anisotropic confinement potentials and results in the breaking of the symmetry between the pristine in-plane $g_{xx}$ and $g_{yy}$ components. We describe the squeezed harmonic potential by $\omega_x\neq \omega_y$ and the tilt by rotating the in-plane coordinates by an angle $\phi_\text{Tilt}$. The resulting potential is then expressed by $V_{xy}(R_z(\phi_\text{Tilt})\textbf{r}_{\parallel})$, where $R_z(\phi_\text{Tilt})$ is a three-dimensional rotation matrix about the z-axis. We now leverage the different responses for different tilt directions to characterize the charge defects.\par 

In Fig.~\ref{fig:sixpanel} we show the variation in g-factor generated by a single charge trap located 30 nm from the origin in the plane of the SiGe/Oxide interface, for different in-plane magnetic field directions and quantum dot tilts. We consider two trap positions, top-left and top-right of the hole qubit, at polar angles $63^\circ$ and $117^\circ$ with respect to the x-axis, with the two values chosen randomly. In this simulation, the planar harmonic confinement produces an anisotropic in-plane g-tensor with its principal axis slightly tilted from the y-axis due to epitaxial strain, making the qubit intrinsically sensitive to trap orientation. This effect, along with the charge disorder, removes the periodicity in the rotation of the $\phi_\text{Tilt}$ angle. These considerations are exemplified in Fig.~\ref{fig:sixpanel}(a), showing the g-tensor $g_0$ subject to charge disorder and strain fluctuations, with the position of the charge fluctuators pictorially represented along with the possible displacement direction. The effect of the dipole of the fluctuator is then considered by evaluating $\delta g=g-g_0$ for the two trap positions and two realizations of $\phi_\text{Tilt}=(0,\pi/2)$. The former is presented in Fig.~\ref{fig:sixpanel}(b) while the latter is reported in Fig.~\ref{fig:sixpanel}(c). \par Our results show that the microscopic response of the g-tensor depends strongly on trap position and displacement. By rotating the quantum dot in the plane, i.e.\ varying its tilt angle, one can tune the contrast between different trap configurations.

\begin{figure}
  \centering
  % ---------- first row ----------
  \includegraphics[width=\linewidth]{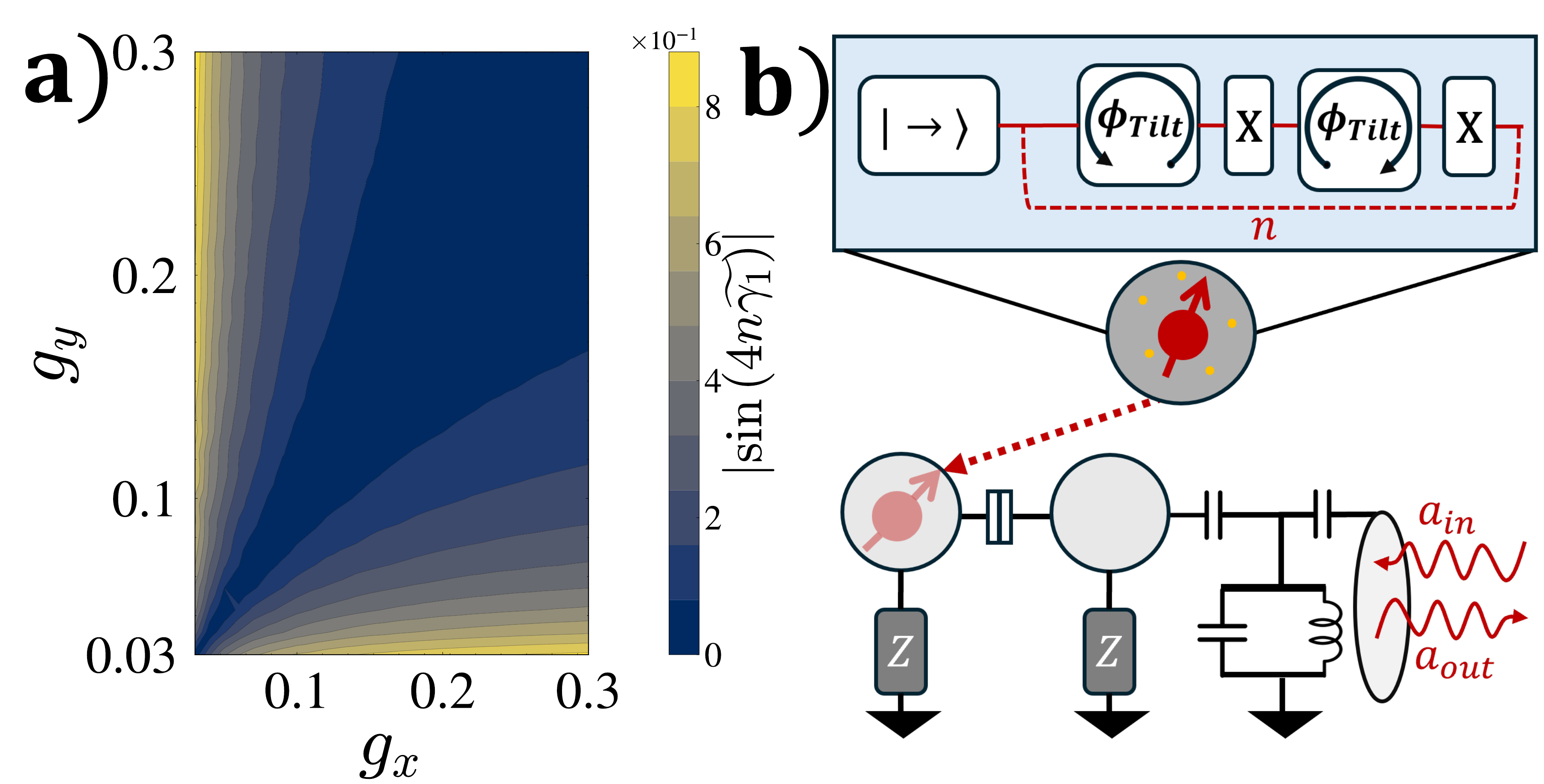}
 \captionsetup{justification=raggedright,singlelinecheck=false}
  \caption{a) Numerical expectation value \( \langle Y \rangle \) after \( n = 10 \) cycles of the Berry phase protocol, plotted as a function of the in-plane g-tensor components \( g_x \) and \( g_y \). b) Sketch of the envisioned architecture for readout of individual g-tensor-component fluctuations. The single spin is prepared in the central dot subject to electrostatic disorder and the TLFs (orange ensemble). Then the tilt-echo protocol is applied $n$ times, where the $\circlearrowright$ represents a clockwise ramp from $0$ to $2 \pi$, and $\circlearrowleft$ is the palindromic ramp of the electrostatic confinement tilt angle. During the phase-accumulation stage, the active TLF is assumed to remain in a fixed metastable state. Finally, the spin is shuttled to the readout zone in the bottom part of the figure, where the flopping mode qubit is shown as two tunnel coupled dots and dispersively coupled to the resonator. }
\label{readout-Figure}
\end{figure}
\par \textit{Readout --- }
For an in-plane magnetic field \(\mathbf B=(B_x,0,0)\), we introduce a geometric tilt-echo protocol that converts a small \(g_{zx}\) component of the \(g\)-tensor into a measurable Berry phase while echoing out the accompanying dynamical phase~\cite{berrydyndecoupl}.
The resulting geometric phase can then be read out either via conventional Pauli spin blockade or dispersively through a coupled microwave cavity using homodyne detection. The protocol, shown in Fig.~\ref{readout-Figure}b, consists of five steps: (i) the hole qubit in a squeezed quantum dot is initialized in an equal superposition of the local eigenstates; (ii) the squeezing axis is rotated adiabatically clockwise through one full cycle; (iii) a calibrated $X$ gate swaps the two local eigenstates; (iv) the squeezing axis is rotated adiabatically counterclockwise through one full cycle; and (v) the final state is measured in the local $y$ basis. We assume that the electrostatic configuration remains fixed during the phase-accumulation stage. Under this quasistatic assumption, the measured Berry phase reflects the small $g_{zx}$ component present during a given accumulation cycle.

For the protocol it is convenient to decompose the full $g$ tensor of a tilted squeezed dot as
\[
\hat g(\phi_{\mathrm{Tilt}})
=
\hat g_{\parallel}(\phi_{\mathrm{Tilt}})
+
\Delta \hat g(\phi_{\mathrm{Tilt}}),
\]
where
\begin{equation}
\hat{g}_{\parallel}(\phi_{\mathrm{Tilt}})
=
\hat{R}_{z}(\phi_{\mathrm{Tilt}})
\,\mathrm{diag}(g_x,\, g_y,\, g_\perp)\,
\hat{R}_{z}(-\phi_{\mathrm{Tilt}})
\label{eq:gparallel}
\end{equation}
is the rotated diagonal background, with $\hat R_z(\phi_{\mathrm{Tilt}})$ the rotation matrix about the $z$ axis. The residual term $\Delta \hat g(\phi_{\mathrm{Tilt}})$ contains all parts of the full tensor not captured by $\hat g_{\parallel}(\phi_{\mathrm{Tilt}})$. It should therefore not be identified with the disorder-induced correction $\delta g$ introduced in the microscopic part. In what follows, $\Delta \hat g(\phi_{\mathrm{Tilt}})$ is treated as a small correction to $\hat g_{\parallel}(\phi_{\mathrm{Tilt}})$. %$\hat{g}_{0}$ contains only even harmonics in $\phi_{\mathrm{Tilt}}$. 

A single in-plane magnetic-field orientation is sufficient for the protocol. For an effective field
\[
\vec{B}_{\mathrm{eff}}=\hat g(\phi_{\mathrm{Tilt}})\vec B
\]
with spherical angles $(\tilde{\theta},\tilde{\phi})$, we choose the local eigenstates as
\begin{align*}
|+\rangle&=
\begin{pmatrix}
\cos\!\tfrac{\tilde{\theta}}{2}\,e^{-i\tilde{\phi}/2}\\[2pt]
\sin\!\tfrac{\tilde{\theta}}{2}\,e^{i\tilde{\phi}/2}
\end{pmatrix},
&
|-\rangle&=
\begin{pmatrix}
-\sin\!\tfrac{\tilde{\theta}}{2}\,e^{-i\tilde{\phi}/2}\\[2pt]
\cos\!\tfrac{\tilde{\theta}}{2}\,e^{i\tilde{\phi}/2}
\end{pmatrix}.
\end{align*}
This gauge choice gives the Berry connections~\cite{xiaoBerryPhaseEffects2010}
$A_{\tilde{\phi}}^{\pm}=i\langle\pm|\partial_{\tilde{\phi}}|\pm\rangle=\pm\tfrac{1}{2}\cos\tilde{\theta}$ and
$A_{\tilde{\theta}}^{\pm}=0$.
In the following, we evaluate all expressions to linear order in $\Delta \hat g(\phi_{\mathrm{Tilt}})$. In particular,
$
\cos\tilde{\theta}\simeq 
\frac{\Delta g_{zx}(\phi_{\mathrm{Tilt}})}
{\sqrt{g_{\parallel,xx}^2(\phi_{\mathrm{Tilt}})+g_{\parallel,yx}^2(\phi_{\mathrm{Tilt}})}}.
$
We now employ a Ramsey-like sequence to read out the small $g_{zx}$ component. Initializing in the equal superposition
$
\frac{|+\rangle_0+|-\rangle_0}{\sqrt{2}},
$
where $|\pm\rangle_0 \equiv |\pm(\phi_{\mathrm{Tilt}}=0)\rangle$ are the local eigenstates of the Hamiltonian at the start of the protocol, we adiabatically ramp the tilt angle $\phi_{\mathrm{Tilt}}$ from $0$ to $2\pi$ by rotating the squeezed potential \footnote{Small quasi-static deviations between this ideal basis and a calibration basis defined for a slightly modified $g$ tensor affect the signal only at higher order; see Sec.~\ref{sec:miscal}.}.
 During this evolution, the state acquires both a geometric (Berry) phase and a dynamical phase, denoted by $\pm\tilde{\gamma}_1$ and $\pm\varepsilon_1$, respectively. To linear order in $\Delta \hat g$, the Berry phase is
\begin{equation}
    \tilde{\gamma}_1
    =
    \frac{1}{2}\int_{0}^{2\pi} d\phi_{\mathrm{Tilt}}\,
    \frac{\Delta g_{zx}(\phi_{\mathrm{Tilt}})}{D(\phi_{\mathrm{Tilt}})}
    \left(
    1-\frac{g_x g_y}{D^2(\phi_{\mathrm{Tilt}})}
    \right),
\end{equation}
where
\begin{equation}
    D(\phi_{\mathrm{Tilt}})
    =
    \sqrt{
    g_x^2\cos^2\phi_{\mathrm{Tilt}}
    +
    g_y^2\sin^2\phi_{\mathrm{Tilt}}
    }.
\end{equation}
After the forward ramp, the state reads
\[
|\psi_{2\pi}\rangle
=
\frac{1}{\sqrt{2}}
\left(
e^{i(\tilde{\gamma}_1+\varepsilon_1)}|+\rangle_0
+
e^{-i(\tilde{\gamma}_1+\varepsilon_1)}|-\rangle_0
\right).
\]

To cancel the dynamical phase, we apply a local $X$ gate \footnote{Small miscalibrations of the fixed calibration basis enter only at higher order; see Sec.~\ref{sec:miscal} of the Supplemental Material.}, defined in the basis $\{|+\rangle_0,|-\rangle_0\}$, which swaps the two states, and then reverse the sweep, $\phi_{\mathrm{Tilt}}:2\pi\rightarrow 0$. For a palindromic sweep $\phi_{\mathrm{Tilt}}(t)$, the dynamical phase cancels exactly, while the geometric phase adds over the two halves of the sequence.

The final state after one full cycle is therefore
\begin{equation}
    |\psi_f\rangle
    =
    \frac{1}{\sqrt{2}}
    \left(
    e^{i2\tilde{\gamma}_1}|-\rangle_0
    +
    e^{-i2\tilde{\gamma}_1}|+\rangle_0
    \right).
\end{equation}
After applying a second $X$ gate and measuring in the local $y$ basis, we obtain
\begin{equation}
    \langle Y \rangle = \sin(4 n \tilde{\gamma}_1)
\end{equation}
after $n$ repetitions of the cycle. The Berry phase vanishes to linear order for an isotropic in-plane $g$ tensor, $g_x=g_y$; details of the derivation are given in the SM ~\cite{SM}.

\par For readout, the spin can be shuttled either to a resonator-coupled flopping-mode double dot~\cite{benitoInputoutputTheorySpinphoton2017,burkardSuperconductorSemiconductorHybridcircuit2020} or to a double dot with an ancilla spin for Pauli spin blockade readout \footnote{To compensate phase evolution during shuttling, a recovery gate is required to correct both for spin dynamics during transport and for the misalignment between the initial and final quantization axes. Imperfect recovery gates may be partially mitigated in post-processing using preceding and succeeding reference measurements~\cite{ademiDistributingEntanglementDistant2025}. Shuttling times can be of order $100\,\mathrm{ns}$~\cite{ademiDistributingEntanglementDistant2025}. }.
Using the input-output formalism, we estimate an SNR of order $1$--$10$ for resonator readout after an integration time of about $100\,\mathrm{ns}$--$1\,\mu\mathrm{s}$ for the parameters assumed in the Supplemental Material~\cite{SM}, while typical Pauli spin blockade readout times are $1$--$20\,\mu\mathrm{s}$~\cite{johnTwodimensional10qubitArray2025a,ademiDistributingEntanglementDistant2025}. The Berry-phase accumulation requires about $10\,\mu\mathrm{s}$ to reach $|\langle Y\rangle|\sim 0.1$ at magnetic fields of order $10\,\mathrm{mT}$, since the tilt rotation must remain slow compared to the Larmor frequency to satisfy the adiabatic condition \footnote{The protocol could be further accelerated using shortcuts to adiabaticity~\cite{glaserTrainingSchrodingerCat2015,theisCounteractingSystemsDiabaticities2018,kochQuantumOptimalControl2022,meinersenUnifyingAdiabaticStatetransfer2025,meinersenUnifyingAdiabaticStatetransfer2025,ventura-meinersenMultilevelSpectralNavigation2026}.}. The total protocol duration therefore remains in the tens-of-microseconds range. Figure~\ref{readout-Figure}a shows the numerical tilt-echo signal for a range of $g_x$ and $g_y$ values. The response is largest for strongly anisotropic in-plane $g$ tensors. An analogous protocol addresses the $g_{zy}$ channel for an in-plane magnetic field $\mathbf B\parallel y$, with the initialization, echo gate, and measurement defined in the corresponding local eigenbasis. Beyond the quasistatic limit, finite-frequency fluctuations modify the signal; a filter-function description is given in the Supplemental Material ~\cite{SM}.

\par To linear order, the Berry-phase signal is sensitive to the out-of-plane tilt of the effective field, i.e.\ to a small $g_{zx}$ contribution on top of the in-plane background. The protocol thus isolates the $g_{zx}$ channel at the level of the geometric phase, but not its microscopic origin: the same signal may arise from electrostatic disorder, non-separability of the confinement potential, or spatially inhomogeneous strain.

\textit{Quantum Fisher  Information --- } We use the quantum Fisher information (QFI) $\mathcal{F}_{\mu\nu}$ to quantify the sensitivity of the qubit state to variations of individual $g$-tensor components~\cite{lambertClassicalQuantumInformation2023,sidhuGeometricPerspectiveQuantum2020}~\footnote{We note, that the QFI provides a lower bound on the covariances of unbiased parameter estimates $\{\hat{x}^\mu\}$ through the Cramer-Rao bound $\text{Cov}[\hat{x}_\mu, \hat{x}_\nu]\geq 1/\mathcal{F}_{\mu\nu}$. To fulfill the Cramer-Rao bound with respect to a given quantity, measurements have to be executed in the eigenbasis of the symmetric logarithmic derivative operator which potentially depends on the estimate parameter itself.}. 
\begin{figure} 
  \centering
  % ---------- first row ----------
  \includegraphics[width=\linewidth]{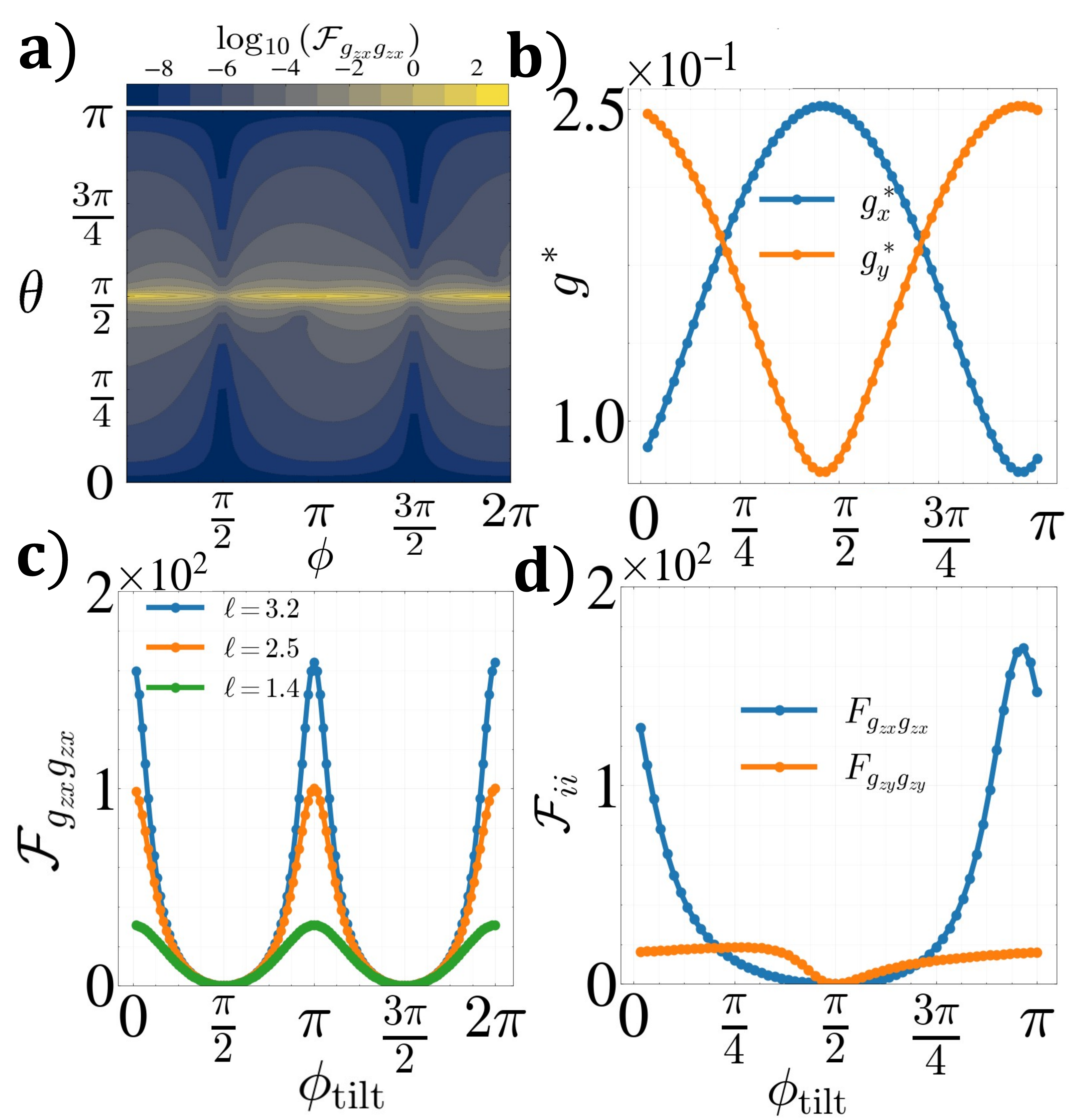}
 \captionsetup{justification=raggedright,singlelinecheck=false}
  \caption{a) Numerical QFI $\mathcal{F}_{g_{zx}g_{zx}}$ in log-scale for the out-of-plane angle between $\theta \in[0,\pi]$ and $\phi \in [0, 2\pi]$. The considered g-tensor corresponds to the one depicted in Fig.~\ref{fig:sixpanel}a). b) The anisotropic planar g-factor is drawn for the magnetic field directions $B=(B_x,0,0)$ and $B=(0,B_y,0)$, for a tilting $\phi_\text{tilt} \in [0, \pi]$. This plot shares the x-axis of panel d). c) Numerical QFI $\mathcal{F}_{g_{zx}g_{zx}}$ for three aspect ratios of in-plane confinement $\ell=L_y/L_x = \sqrt{\omega_x/\omega_y}$. d) The $\mathcal{F}_{g_{zx}g_{zx}}$ and $\mathcal{F}_{g_{zy}g_{zy}}$ corresponding to the anisotropic planar g-factor of panel b).
  }
\label{QFI-Figure}
\end{figure}
The same geometric mechanism underlies both the readout protocol, which is sensitive only to the $g_{zx}$ component, and the enhanced QFI for the $g_{zx}$ channel relative to the $g_{xx}$ channel, as illustrated in Fig.~\ref{QFI-Figure}. For 
\begin{equation}
    H_\text{QFI} = \mu_B B_x \left(g_{xx} \sigma_x + g_{yx} \sigma_y + g_{zx} \sigma_z \right),
\end{equation}
the reference tensor $\hat g_{\parallel}$ introduced in Eq.~\eqref{eq:gparallel} has $g_{zx}=0$ by construction, so its effective field lies in the plane.  Since both the QFI and the Berry-phase response are determined by changes in the eigenstates, only the component of a $g$-tensor variation perpendicular to the effective field contributes. A purely longitudinal variation, which only rescales the field magnitude, does not change the eigenstates. For the reference tensor $\hat g_{\parallel}$, the effective field lies in the plane, so a small residual $g_{zx}$ produces a transverse perturbation already at linear order. This enhances
\begin{equation}
    \mathcal{F}_{g_{zx}g_{zx}}
    =
    \frac{g_{xx}^2 + g_{yx}^2}{(g_{xx}^2 + g_{yx}^2 + g_{zx}^2)^2}
    \xrightarrow[g_{zx}\rightarrow 0]{}
    \frac{1}{g_{xx}^2 + g_{yx}^2},
\end{equation}
whereas
\begin{equation}
    \mathcal{F}_{g_{xx}g_{xx}}
    =
    \frac{g_{yx}^2 + g_{zx}^2}{(g_{xx}^2 + g_{yx}^2 + g_{zx}^2)^2}
    \xrightarrow[g_{zx}\rightarrow 0]{}
    \frac{g_{yx}^2}{(g_{xx}^2 + g_{yx}^2)^2}.
\end{equation}
Thus, at $g_{zx}=0$, one has $\mathcal{F}_{g_{zx}g_{zx}}>\mathcal{F}_{g_{xx}g_{xx}}$: varying $g_{zx}$ is purely transverse, whereas varying $g_{xx}$ has a longitudinal component and therefore changes the eigenstates less efficiently. The Berry-phase response is more restrictive.
For the $\hat{g}_{\parallel}$ considered here Eq.~\eqref{eq:gparallel}, the unperturbed effective-field trajectory encloses no solid angle, so the Berry phase vanishes. To first order, $\Delta g_{xx}$ and $\Delta g_{yx}$ only deform the unperturbed trajectory within its plane and therefore do not modify the Berry phase, whereas $\Delta g_{zx}$ lifts it out of plane and produces a nonzero geometric response. In-plane anisotropy is essential in both cases, but for different reasons: for the QFI, it reduces the magnitude of the in-plane effective field and thereby enhances the response to $\Delta g_{zx}$; for the Berry phase, it prevents the unperturbed trajectory from collapsing to a point, which would eliminate the response in the linear order.
 \newline \indent In Fig.~\ref{QFI-Figure}a, we show $\mathcal{F}_{g_{zx}g_{zx}}$ as a function of the magnetic-field angles $(\theta,\phi)$. The largest values occur for in-plane magnetic fields, in agreement with the analytical discussion in the SM ~\cite{SM}. For fields aligned along the $x$ axis, $\mathcal{F}_{g_{zx}g_{zx}}$ is strongly enhanced; see Fig.~\ref{QFI-Figure}b--d. Figure~\ref{QFI-Figure}c and Fig.~\ref{readout-Figure}a show the same dependence on in-plane anisotropy: increasing the squeezing aspect ratio enhances the sensitivity to the $g_{zx}$ channel.

\textit{Conclusion --- }In this work, we have shown that germanium hole qubits exhibit a directional response to nearby two-level fluctuators through the resulting variations of the $g$ tensor. By analyzing the response for different trap locations and different tilts of the harmonic confinement, we identified clearly distinguishable signatures. We further introduced a geometric readout protocol that converts a small out-of-plane $g$-tensor component into a measurable Berry phase while suppressing the accompanying dynamical phase, on a total timescale of tens of microseconds. Finally, by computing the quantum Fisher information, we identified in-plane magnetic field directions and anisotropy regimes that enhance the sensitivity to variations of individual $g$-tensor components. Together, these results establish a concrete route to probing disorder-induced $g$-tensor variations in germanium hole spin qubits.

\textit{Acknowledgments --- }We thank all members of the Bosco, Rimbach-Russ, Veldhorst, and Vandersypen group for valuable feedback. We acknowledge helpful discussions with G. Katsaros, L. Cywinski and M. Pham Nguyen. We further thank G. Katsaros for providing information about the experimental measurements. M.R.-R. and E.V. acknowledge support from the Dutch Research Council (NWO) under Award Number Vidi TTW 22204. This research was further supported by the EU through the H2024 QLSI2 project and partly sponsored by the Army Research Office under Award Number: W911NF-23-1-0110. The views and conclusions contained in this document are those of the authors and should not be interpreted as representing the official policies, either expressed or implied, of the Army Research Office or the U.S. Government. The U.S. Government is authorized to reproduce and distribute reprints for Government purposes notwithstanding any copyright notation herein.

\textit{Author contributions --- } E.V. performed the microscopic simulations. C.V.M. and T.V. derived the expressions for the quantum Fisher information. T.V. developed and simulated the readout protocol and extended the analysis to finite-frequency noise via the filter-function formalism. E.V. and M.J. developed the simulation software. M.R.-R. conceived the project. E.V., C.V.M., T.V., and M.R.-R. wrote the manuscript, with input from M.J. and S.B. \par
\textit{Data Availability.-- }The data that support the findings of
this article are openly available~\cite{Zenodo}

\bibliographystyle{apsrev4-1}
\bibliography{references}

\clearpage 

\widetext
\beginsupplement

\begin{center}
  {\Large \textbf{Supplementary Material}}\\[4pt]
  {\large \textbf{Probing Electrostatic Disorder via g-tensor Geometry}}\\[6pt]
  Edmondo Valvo, Christian Ventura-Meinersen, Michèle Jakob,\\
  Stefano Bosco, Tereza Vakhtel, Maximilian Rimbach-Russ\\[2pt]
  (Dated: \today)
\end{center}

\section*{Abstract}
In this supplementary material we look at the effect of a larger charge ensemble on the results derived in the main text. Furthermore, we derive thoroughly the tilt-echo protocol and the readout procedure.

\makeatletter
\renewcommand{\citenumfont}[1]{S#1}
\makeatother
\section{Details of microscopic model}
In this section we present the details of the microscopic model used for the simulations in the main text. The band structure is obtained via the $4\times4$ Luttinger-Kohn hamiltonian 
\begin{equation}\label{LK4by4}
\begin{aligned}
     H_{LK}= \frac{\hbar^2}{2 m_0}& \Big[(\gamma_1 +\frac{5 \gamma_2}{2}) \frac{\textbf{k}^2}{2} - \gamma_2  \sum_{i}(k_i^2 J_i^2) - 2 \gamma_3 \sum_{i<j} \{k_i,k_j\}\{J_i,J_j\}\Big] ,
\end{aligned}
\end{equation}
where $\gamma_1=13.38,\gamma_2=4.24,\gamma_3=5.69$ are the Luttinger parameters. 
Furthermore, the strain is described by the Bir-Pikus Hamiltonian which is given by\begin{equation}
 H_{BP}=\begin{bmatrix}
 P_\epsilon +Q_\epsilon  & 0 & -S_\epsilon  & R_\epsilon  \\
 0 & P_\epsilon +Q_\epsilon  & R_\epsilon ^* & S_\epsilon ^* \\
 -S_\epsilon ^* & R_\epsilon  & P_\epsilon -Q_\epsilon  & 0 \\
 R_\epsilon ^* & S_\epsilon  & 0 & P_\epsilon -Q_\epsilon  \\
 \end{bmatrix}   
\end{equation}
where $P_{\epsilon}=-a_v (\epsilon_{xx} + \epsilon_{yy} + \epsilon_{zz})$, $Q_\epsilon=-b_v/2 (\epsilon_{xx} + \epsilon_{yy} -2 \epsilon_{zz})$, $R_{\epsilon}=\frac{1}{2} \sqrt{3} b (\epsilon_{xx}-\epsilon_{yy})-i d \epsilon_{xy}$, $S_\epsilon=-d (\epsilon_{xz}-i \epsilon_{yz})$ and the microscopic values being $a_v=-2$ eV,  $b_v=-2.3$ eV and $d_v=-6$ eV ~\cite{Terrazos2021}. To simulate the fluctuations of the strain pattern coming from misfit dislocations typical in strained Ge heterostructures we set $\epsilon_{xx}=\epsilon_{yy}=-0.61\%$, $\epsilon_{zz}=0.1\%$, $\epsilon_{xz}=0.01\%$, $\epsilon_{yz}=0.01\%$ and $\epsilon_{xy}=0.001\%$.
The response of the spin qubit to a magnetic field, assuming linear dependence, can be described as $H_B=H_{\text{Zeeman}}+ 2\mu_B [\gamma_3\{\textbf{A}\cdot\textbf{J},\textbf{k}\cdot \textbf{J}\}+(\gamma_2-\gamma_3)\{\textbf{A},\textbf{k}\}\cdot\textbf{J}^2]$ where 
$ H_{\text{Zeeman}}= 2 \mu_B \kappa \textbf{J}\cdot \textbf{B} + 2 \mu_B q \sum_{i}(J_i^3 B_i)$.
Finally, the z-direction potential is a piecewise function for the SiGe buffers and the Ge well. More precisely, we set $-eF_z z$ inside the quantum well $(-L<z<0)$ and $-eF_z z + U_{\text{SiGe}}$ otherwise. By including a finite barrier, the $z$-direction solution to the Schroedinger equation takes the form of a linear combination of Airy functions ~\cite{Wang2024Modelling}. To circumvent the computational complexity of these peculiar functions and drastically improve the computational cost of Coulomb integrals, the exact solution is approximated with a linear combination of Gaussian functions $\phi_z(z)=\sum_ia_i G_i(z)$, where  $G_i(z)=e^{-\frac{(z-\mu_i)^2}{\sigma_i^2}}$. Our Gaussian wavefunction basis automatically considers a finite penetration into the SiGe buffer layers. 
Furthermore, in the model we place an ensemble of charges at the SiGe/Oxide interface mimicking electrostatic disorder often found in realistic experimental devices. 
These charges are distributed respecting a surface density of $\unit[10^{10} ]{cm^{-2}}$ ~\cite{Kpa2023} and, unlike TLFs, they are full Coulomb potentials that are screened by the top gate:
\begin{equation}
    V=\sum_i - \frac{F_c}{|\textbf{r}_i|}  + \frac{F_c}{|\textbf{r}_i+\textbf{r}_m|}
\end{equation}

\section{Robustness against additional fluctuators}
\begin{figure}[h]
    \centering
    \includegraphics[width=0.8\linewidth]{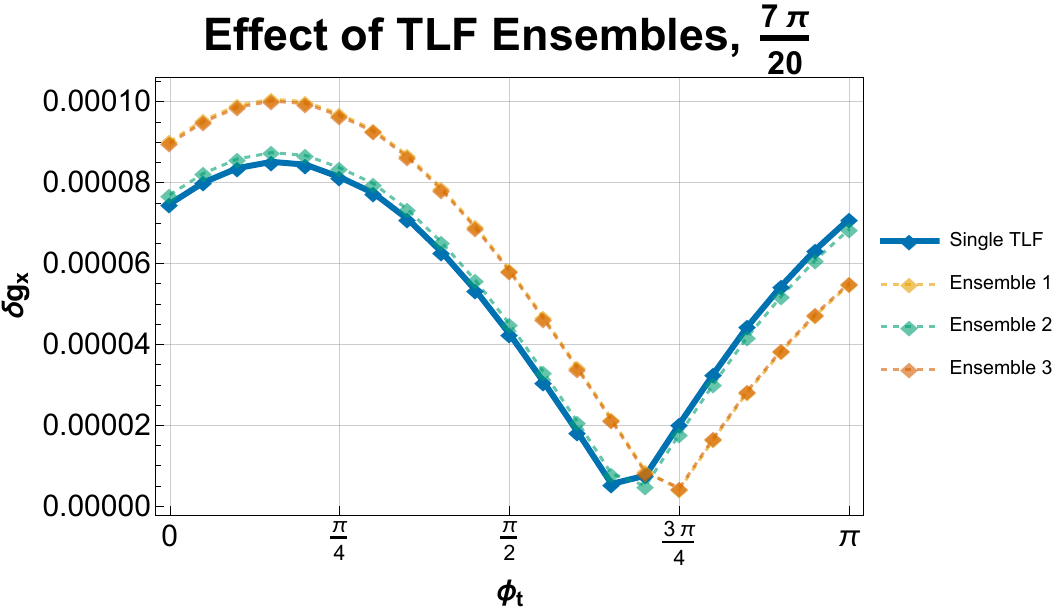}
    \caption{Variation of $\delta g_x$ due to a bath of TLFs with the strongly coupled TLF placed at $63^\circ$ from the x-axis.}
    \label{fig:ensembles 7Pi20}
\end{figure}

A fundamental assumption in this study is the presence of an individual charge fluctuator and therefore a relatively pristine material. This choice was dictated by the observation that in typical experimental devices a single TLF couples significantly to the Ge hole qubit. However, it is reasonable to assume that a realistic device presents an ensemble of charge fluctuators sparsely distributed with an estimated density of $\unit[10^{10} ]{cm^{-2}}$ ~\cite{Kpa2023}. To take into account this effect we have performed simulations with including a bath of charge fluctuators modeled according to the following potential:
\begin{equation}
    V_{c,\text{Ensemble}}=\sum_i\frac{F_c}{|\textbf{r}_i+\delta\textbf{r}_i|} - \frac{F_c}{|\textbf{r}_i|} - \frac{F_c}{|\textbf{r}_i+\delta\textbf{r}_i + \textbf{r}_m|} + \frac{F_c}{|\textbf{r}_i+\textbf{r}_m|}
\end{equation}
We place uniformly a number of charge traps around the quantum dot to match the charge density measurements. The result for the case of a strongly coupled TLF at $63^\circ$ from the x-axis, is reported in \cref{fig:ensembles 7Pi20} for a handful of realizations.
\begin{figure}[h]
    \centering
    \includegraphics[width=0.8\linewidth]{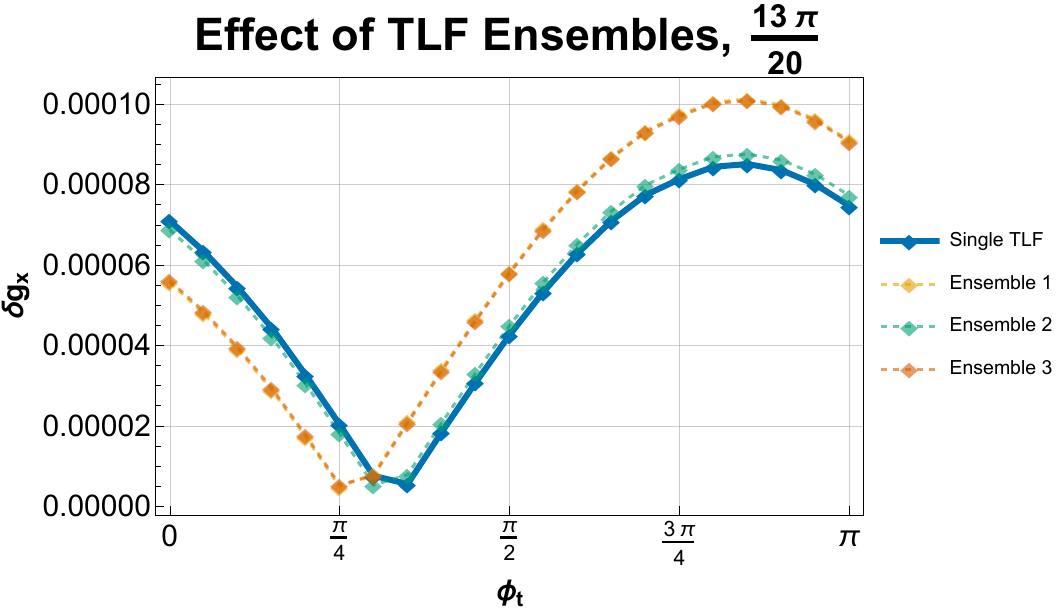}
    \caption{Variation of $\delta g_x$ due to a bath of TLFs with the strongly coupled TLF placed at $117^\circ$ from the x-axis. }
    \label{fig:ensembles 13Pi20}
\end{figure}
\begin{figure}[h]
    \centering
    \includegraphics[width=0.8\linewidth]{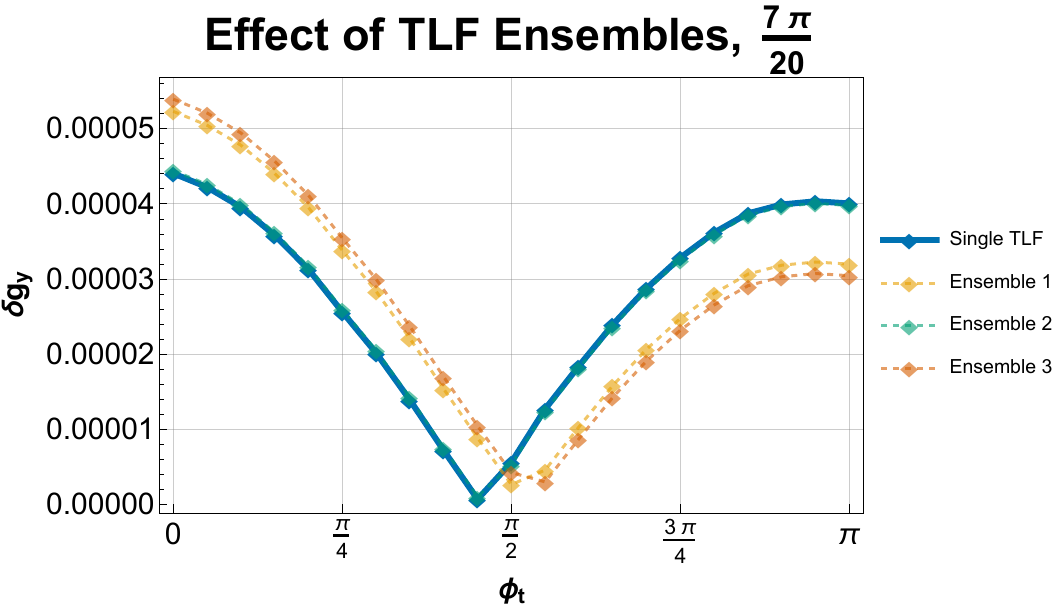}
    \caption{Variation of $\delta g_y$ due to a bath of TLFs with the strongly coupled TLF placed at $63^\circ$ from the x-axis. }
    \label{fig:ensembles 7Pi20 y}
\end{figure}
\begin{figure}[h]
    \centering
    \includegraphics[width=0.8\linewidth]{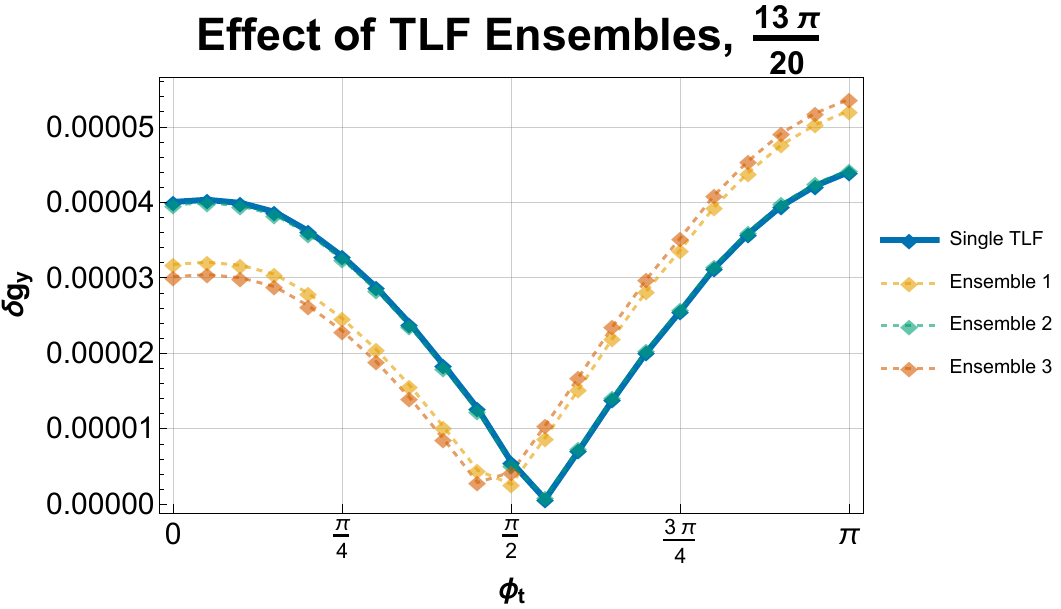}
    \caption{Variation of $\delta g_y$ due to a bath of TLFs with the strongly coupled TLF placed at $117^\circ$ from the x-axis. }
    \label{fig:ensembles 13Pi20 y}
\end{figure}

These simulations show that depending on the coupling strength of the ensemble, the insight from the main work can either be preserved or slightly modified. In particular, it is apparent that the overall symmetry of the $g_x$ response is preserved for all the considered ensembles, with the traps on either side of the dot showing minima symmetrically with respect to a $\frac{\pi}{2}$ tilt. However, when looking at the variation of $g_y$, charge ensembles 1 and 3 shift the original response beyond the $\phi_t=\frac{\pi}{2}$ symmetry point, making the detection more challenging. In conclusion this analysis confirms the initial assumption that relies on a relatively clean sample that presents a single dominant fluctuator. It is worth noting that in these simulations, to properly assess the effect of other dipoles, we remove strain fluctuations and overall electrostatic disorder that is not due to TLFs.

\newpage 

\section{Quantum Fisher Information}
\label{app: QFI}

\subsection{Pedagogical introduction to Quantum Fisher information}
In the main text, we use the relationship between the eigenvalues and eigenstates of a Hamiltonian and the Fisher information to address and examine parameter regimes that allow for a tighter bound on the covariance of parameter estimates. Here we derive this relationship. We start with a pure state $\rho(x)=\ketbra{\psi(x)}$ that is defined by some parameters $x=x^\mu$. We can define tangent vectors $t_\mu$ along these parameters as follows
\begin{align}
    t_\mu(x)=\partial_\mu \hat{\rho}(x)=\ketbra{\partial_\mu \psi}{\psi}+\ketbra{\psi}{\partial_\mu \psi}.
\end{align}
We can define the Fisher information (quantum metric tensor) as the Killing form on the tangent space $T_{\hat{\rho}} P(\mathcal{H})$, where $P(\mathcal{H})=\mathcal{H}/U(1)$
\begin{align}
    \mathcal{F}_{\mu \nu}\propto\frac{1}{2}\tr(t_\mu t_\nu)
    =\Re\Big[\braket{\partial_\mu \psi}{\partial_\nu \psi}\Big]+\braket{\partial_\mu \psi}{\psi}\braket{\psi}{\partial_\nu \psi}.
\end{align}

\begin{figure}
    \centering
    \includegraphics[width=\linewidth]{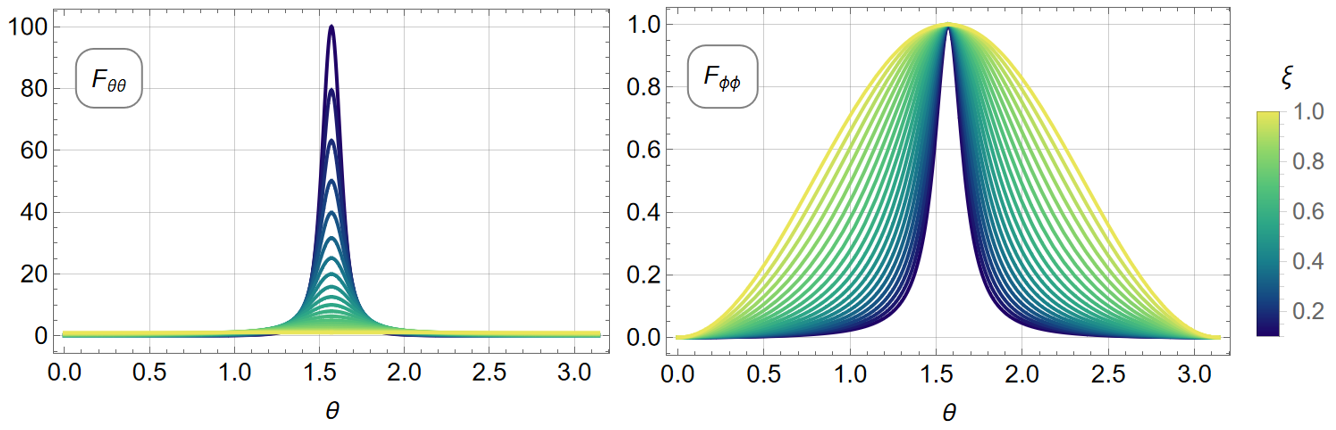}
    \caption{Quantum Fisher information as a function of out-of-plane magnetic field direction $\theta$ for different values of anisotropy $\xi$. The maxima are always found in the in-plane direction $\theta=\pi/2$.}
    \label{fig: supp QFI}
\end{figure}

In this section, we compute the quantum Fisher information for a non-isotropic g-tensor. We aim to identify the sensitivity hotspots as a function of magnetic field angle. To find analytic solutions, we approximate the microscopic Hamiltonian by an effective two-level Hamiltonian capturing the effects of the g-tensor as
\begin{align}
    \hat{H}_\text{eff}=\frac{1}{2}\mu_B B\, \Vec{\sigma}\cdot \Big(g\hat{b}(\theta,\phi)\Big),
\end{align}
where we write $\hat{b}(\theta,\phi)=\Vec{B}/B$, with $B=|\Vec{B}|$, as the unit vector that depends on the polar ($\phi$) and azimuthal ($\theta$) angle, respectively. We choose a g-tensor with diagonal components $g=\text{diag}(g_\parallel,g_\parallel,g_\perp)$. The QFI $\mathcal{F}_{\mu\nu}$ is defined as
\begin{align}
    \mathcal{F}_{\mu\nu}=4\Re\sum_{m\neq n} \frac{\mel{\psi_n}{\partial_\mu \hat{H} }{\psi_m}\mel{\psi_m}{\partial_\nu \hat{H} }{\psi_n}}{(E_m-E_n)^2},    
\end{align}
where $\{E_n,\ket{\psi_n}\}_{n=0}^{d-1}$ is the set of eigenvalues and eigenvectors of the Hamiltonian and $\partial_\mu=\partial/\partial x^\mu$ is the partial derivative with respect to some parameter $x^\mu$; in our case $x^\mu=\{\theta,\phi\}$, which constitute the magnetic field angles. As the QFI is invariant under local rescaling of the Hamiltonian, i.e., $\hat{H}(x)\to \Omega(x)\hat{H}(x)$ for some parameter-dependent scaling factor $\Omega(x)$ ~\cite{ventura-meinersenQuantumGeometricProtocols2025,meinersenUnifyingAdiabaticStatetransfer2025}, we find
\begin{align}
    \hat{H}_\text{eff}=\mqty(\cos \theta & \xi\, e^{-i\phi}\sin \theta\\
    \xi\, e^{i\phi}\sin \theta & -\cos \theta),
\end{align}
where we defined $\xi\equiv g_\parallel/g_\perp$ as the degree of anisotropy. The QFI with respect to $\{\theta,\phi\}$ reads 
\begin{align}
    \mathcal{F}_{\theta \theta}&=\frac{\xi^2\sec^4 \theta}{(1+\xi^2\tan^2\theta)^2} & \mathcal{F}_{\phi\phi}&=\left(1+\frac{\cot^2\theta}{\xi^2}\right)^{-1}.
\end{align}
The off-diagonal components $\mathcal{F}_{\theta\phi}{=}\mathcal{F}_{\phi\theta}{=}0$. In the fully isotropic limit $\xi\to 1$, one recovers the standard Bloch sphere components $\mathcal{F}_{\theta \theta}\,{=}\,1$ and $\mathcal{F}_{\phi\phi}\,{=}\,\sin^2\theta$. In the general case of a g-tensor anisotropic in-plane and arbitrary tilted and $\xi< 1$, we find that the maximum for all components of the QFI is found at $\theta=\pi/2$, which corresponds to the in-plane direction. Particularly, in the in-plane limit we find $\lim_{\theta\to\pi/2}\mathcal{F}_{\theta\theta}=1/\xi^2$, which demonstrates a big response for strong out-of-plane anisotropy $\xi\ll 1$. In Figure~\ref{fig: supp QFI}, we illustrate the two components for different values of anisotropy $\xi$.

\subsection{Sensitivity to squeezing aspect ratio}
\begin{figure}[h]
    \centering
    \includegraphics[width=0.5\linewidth]{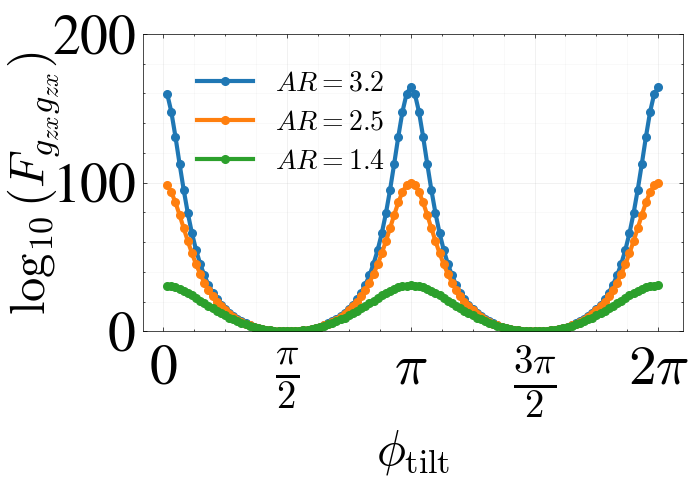}
    \caption{Quantum Fisher information as a function of the planar tilt angle $\phi_\text{Tilt}$ for different values of squeezing aspect ratio (AR). The maxima are always found when the g-factor is at its minimum.}
    \label{fig: AR effect}
\end{figure}

To exemplify the sensitivity of the QFI against the geometry of the planar g-tensor we plot the $\mathcal{F}_{g_{zx}g_{zx}}$ for some simplified g-tensor without off-diagonal shear term components, squeezed along x and with aspect ratios AR=$\sqrt{ \frac{\omega_x}{\omega_y}}= (3.2, 2.5, 1.4)$. The results are shown in Fig.~\ref{fig: AR effect}.
This particular detail gives geometrical insight to the results of the readout protocol of Fig.~\ref{QFI-Figure}(c). In particular, we can observe the largest response when the g-tensor anisotropy is largest, mirroring the results of the QFI analysis.

\subsection{Sensitivity of $g_\text{zx}$}
\begin{figure}[h]
    \centering
    \includegraphics[width=0.8\linewidth]{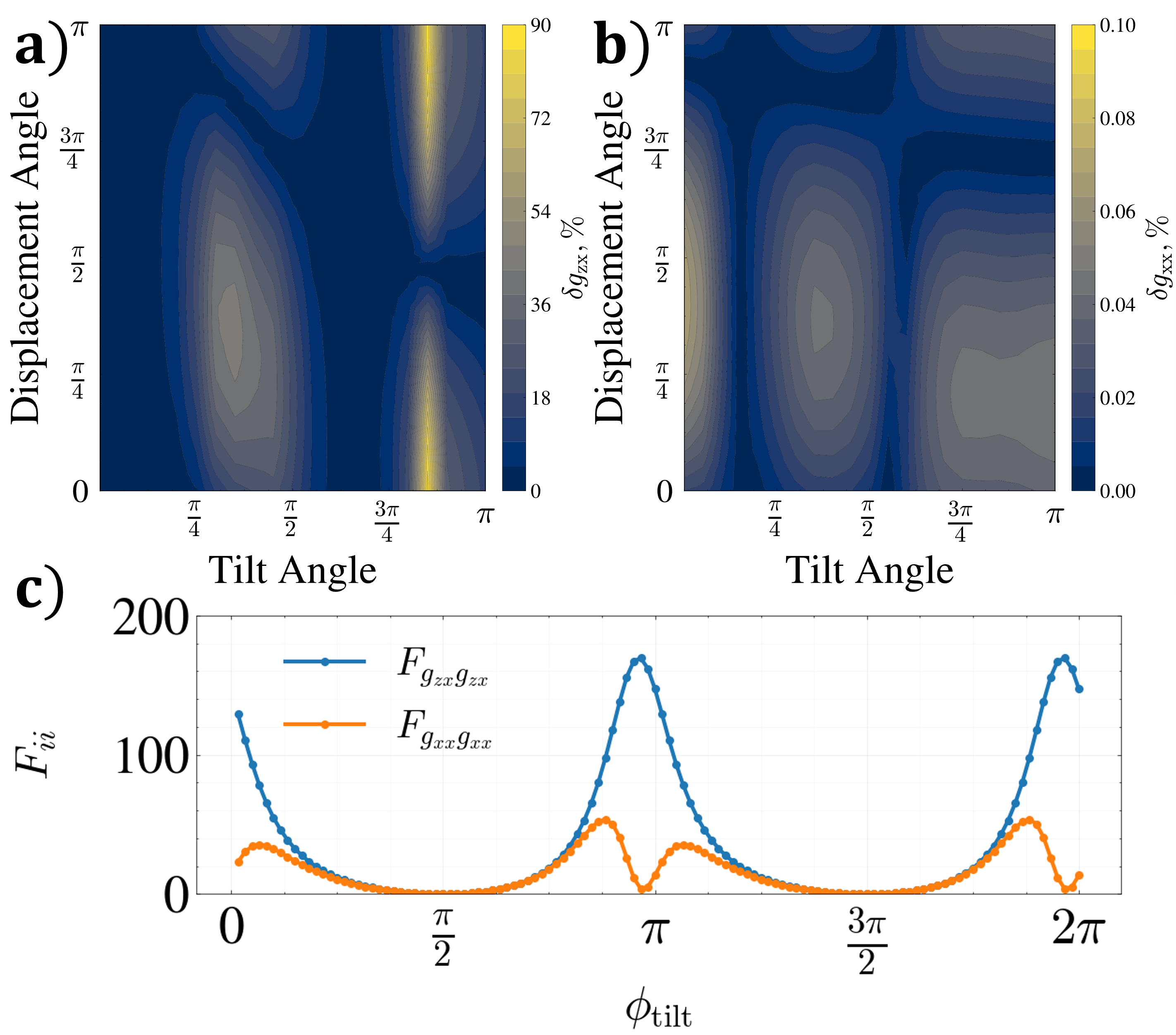}
    \caption{Relative variation of the g-tensor components $g_\text{zx}$ and $g_\text{xx}$ in a) and b) respectively. c) reports the QFI with respect to the principal component $g_\text{xx}$ and the off-diagonal $g_\text{zx}$, similarly to the relative variation $\delta g$ the QFI shows a significantly larger response for the shear term. }
    \label{fig: relative variation}
\end{figure}

To illustrate further the choice of $g_\text{zx}$ as the parameter to readout we simulated the relative variation of the g-tensor components $g_\text{zx}$ and $g_\text{xx}$ when a TLF is present, formally $\delta g_{ij} = \frac{g_{N,ij}}{g_{0,ij}}*100$. The maps are reported in Fig.~\ref{fig: relative variation} and show a significantly larger response for the off-diagonal shear component $g_\text{zx}$, reaching approximately 80 \% while the principal component $g_\text{xx}$ is only limited to 0.08 \%. This is further confirmed by the QFI comparison in Fig.~\ref{fig: relative variation} c) with $\mathcal{F}_{g_\text{zx}}$ at least an order of magnitude improvement over $\mathcal{F}_{g_\text{xx}}$.

\newpage

\section{Derivation of the tilt-echo protocol}
\label{app:berry}

Our protocol follows closely the idea described in \cite{berrydyndecoupl}. We consider the Hamiltonian
\begin{equation}\label{hamilt_prot}
H = H_{0} + \Delta H,
\end{equation}
with
\begin{equation}
H_{0} = \frac{\mu_{B}}{2}\,\hat{\boldsymbol{\sigma}}\cdot \hat g_{\parallel}\mathbf{B},
\qquad \mathbf{B} = (B_{x},0,0),
\end{equation}
and
\begin{equation}
\hat g_{\parallel}(\phi_{\mathrm{Tilt}}) =
\hat{R}(\phi_{\mathrm{Tilt}})
\begin{pmatrix}
g_{x} & 0 & 0 \\
0 & g_{y} & 0 \\
0 & 0 & g_{\perp}
\end{pmatrix}
\hat{R}(-\phi_{\mathrm{Tilt}}),
\end{equation}
where $\hat{R}$ is the rotation matrix in the $x$-$y$ plane. The perturbation
\begin{equation}
\Delta H = \frac{\mu_{B}}{2}\,\hat{\boldsymbol{\sigma}}\cdot\Delta \mathbf{g}\,\mathbf{B},
\end{equation}
contains both even and odd harmonics of $\phi_{\mathrm{Tilt}}$:
\begin{align}
\Delta g_{ix}(\phi_{\mathrm{Tilt}})
=
\sum_{m=0}^{\infty} \Delta g_{ix}^{m}
\cos\!\bigl(m\phi_{\mathrm{Tilt}}+\alpha_{mix}\bigr).
\end{align}

For a spin in an effective magnetic field
\begin{equation}
\mathbf{B}_{eff} = |B|\bigl(\sin\tilde{\theta}\cos\tilde{\phi},
\sin\tilde{\theta}\sin\tilde{\phi},
\cos\tilde{\theta}\bigr),
\end{equation}
we choose the gauge
\begin{equation}\label{gauge}
|+\rangle=
\begin{pmatrix}
\cos(\tilde{\theta}/2)e^{-i\tilde{\phi}/2}\\
\sin(\tilde{\theta}/2)e^{+i\tilde{\phi}/2}
\end{pmatrix},
\qquad
|-\rangle=
\begin{pmatrix}
-\sin(\tilde{\theta}/2)e^{-i\tilde{\phi}/2}\\
\cos(\tilde{\theta}/2)e^{+i\tilde{\phi}/2}
\end{pmatrix},
\end{equation}
giving the Berry connection
\begin{equation}
A_{\tilde{\phi},\pm}
= i\langle \pm| \partial_{\tilde{\phi}}| \pm \rangle
= \pm \frac{\cos \tilde{\theta}}{2}.
\end{equation}
To linear order in $\Delta \hat g$,
\begin{equation}
\cos\tilde{\theta} \approx
\frac{\Delta g_{zx}(\phi_{\mathrm{Tilt}})}{
\sqrt{g_{\parallel,xx}^{2}(\phi_{\mathrm{Tilt}})
+
g_{\parallel,yx}^{2}(\phi_{\mathrm{Tilt}})}} \ll 1.
\end{equation}
The spin azimuthal angle $\tilde{\phi}(\phi_{\mathrm{Tilt}})$ satisfies
\begin{equation}
\tilde{\phi}(\phi_{\mathrm{Tilt}})
=
\arctan\frac{g_{\parallel,yx}(\phi_{\mathrm{Tilt}})}{g_{\parallel,xx}(\phi_{\mathrm{Tilt}})},
\end{equation}
so, to zeroth order in $\Delta \mathbf{g}$,
\begin{equation}
\frac{\partial\tilde{\phi}}{\partial \phi_{\mathrm{Tilt}}}
=
1-\frac{g_{x}g_{y}}{
g_{x}^{2}\cos^{2}\phi_{\mathrm{Tilt}}
+
g_{y}^{2}\sin^{2}\phi_{\mathrm{Tilt}}}.
\end{equation}

Thus the Berry phase is
\begin{equation}
\tilde{\gamma}_{1}
=
\pm\!\int_{\phi_{i}}^{\phi_{f}}
d\phi_{\mathrm{Tilt}}\,
\frac{\Delta g_{zx}(\phi_{\mathrm{Tilt}})}{
2\sqrt{g_{x}^{2}\cos^{2}\phi_{\mathrm{Tilt}}
+
g_{y}^{2}\sin^{2}\phi_{\mathrm{Tilt}}}}
\left(
1-\frac{g_{x}g_{y}}{
g_{x}^{2}\cos^{2}\phi_{\mathrm{Tilt}}
+
g_{y}^{2}\sin^{2}\phi_{\mathrm{Tilt}}}
\right),
\label{eq:gamma}
\end{equation}
where $+$ and $-$ stand for the instantaneous excited and ground state, respectively. For $g_x=g_y$, Eq.~\eqref{eq:gamma} vanishes to linear order.

The dynamical phase accumulated during a $2\pi$ sweep of $\phi_{\mathrm{Tilt}}(t)$ is, including only corrections to linear order in $\|\Delta \hat g\|$,
\begin{align}\label{dyn_phase}
\varepsilon(\phi_f,\phi_i)\approx
&\mu_{B}B_{x}
\int_{\phi_{i}}^{\phi_{f}}\frac{d\phi_{\mathrm{Tilt}}}{2\dot{\phi}_{\mathrm{Tilt}}}\,
\sqrt{
g_{\parallel,xx}^{2}(\phi_{\mathrm{Tilt}})
+
g_{\parallel,yx}^{2}(\phi_{\mathrm{Tilt}})}
\nonumber\\[-2pt]
&\times
\left[
1+
\frac{
g_{\parallel,xx}\Delta g_{xx}(\phi_{\mathrm{Tilt}})
+
g_{\parallel,yx}\Delta g_{yx}(\phi_{\mathrm{Tilt}})
}{
(g_{\parallel,xx})^{2}
+
(g_{\parallel,yx})^{2}}
\right].
\end{align}

\subsection{Protocol for geometric-phase accumulation}

In this protocol, $\phi_{\mathrm{Tilt}}$ is shifted adiabatically. For simplicity, we assume that the $X$ and $Y$ rotations at $\phi_{\mathrm{Tilt}}=0$ are calibrated such that
\begin{equation}
X=\begin{pmatrix} 0&1 \\ 1 & 0 \end{pmatrix}
\end{equation}
in the $|\pm\rangle$ basis. However, it is by no means necessary that the gate has exactly this form in this basis and the results remain the same, as long as, up to corrections linear in $\|\Delta \mathbf{g}\|$, it is orthogonal to $Z$ in the local eigenbasis and the same $X$ and $Y$ are used consistently throughout the protocol.

\begin{enumerate}
\item Prepare the initial superposition by applying an $e^{i\frac{\pi Y}{4}}$ pulse to the ground state:
\begin{equation}
|i\rangle=\frac{|+\rangle_{0}+|-\rangle_{0}}{\sqrt{2}}.
\end{equation}

\item Sweep $\phi_{\mathrm{Tilt}}:0\rightarrow 2\pi$:
\begin{equation}
|\psi_{2\pi}\rangle=
\frac{
e^{+i(\tilde{\gamma}_{1}+\varepsilon_f)}|+\rangle_{0}
+
e^{-i(\tilde{\gamma}_{1}+\varepsilon_f)}|-\rangle_{0}
}{\sqrt{2}},
\end{equation}
where $\varepsilon_f$ is the dynamical phase in Eq.~\eqref{dyn_phase} accumulated when sweeping the tilt angle forward.

\item Apply a local $X$ gate:
\begin{equation}
|\psi_{2\pi}\rangle\to
\frac{
e^{+i(\tilde{\gamma}_{1}+\varepsilon_f)}|-\rangle_{0}
+
e^{-i(\tilde{\gamma}_{1}+\varepsilon_f)}|+\rangle_{0}
}{\sqrt{2}}.
\end{equation}

\item Sweep $\phi_{\mathrm{Tilt}}:2\pi\rightarrow0$:
\begin{equation}
|\psi_{0}\rangle=
\frac{
e^{+i(2\tilde{\gamma}_{1}+\varepsilon_f-\varepsilon_{b})}|-\rangle_{0}
+
e^{-i(2\tilde{\gamma}_{1}+\varepsilon_f-\varepsilon_{b})}|+\rangle_{0}
}{\sqrt{2}},
\end{equation}
where $\varepsilon_b$ is the dynamical phase accumulated when sweeping the tilt angle backwards.

\item Apply another $X$ and measure (this will be measured via the homodyne detection technique in the cQED setup discussed in the main text):
\begin{equation}
\langle Y \rangle = \sin\!\left(4\tilde{\gamma}_{1}+2(\varepsilon_{f}-\varepsilon_{b})\right).
\end{equation}
\end{enumerate}

This expectation value can be made entirely geometric-phase dependent if one ensures that \(\varepsilon_{f} = \varepsilon_{b}\). This condition can be satisfied by making the backward sweep the time-reversed version of the forward one, i.e.\ by applying the transformation \(t \rightarrow \mathcal{T} - t\), where \(\mathcal{T}\) is the total sweep duration for each direction.

Repeating this sequence $n$ times yields $\sin(4n\tilde{\gamma}_{1})$.

All odd-harmonic terms in $\Delta\mathbf{g}$ integrate to zero in Eq.~\eqref{eq:gamma}. The geometric phase is therefore determined entirely by the even-harmonic components:
\begin{equation}
\tilde{\gamma}_{1}=\sum_{n} \tilde{\gamma}^{n}_{1}.
\end{equation}
\begin{equation}
2\tilde{\gamma}^{2n}_{1}
=
\Delta g_{zx}^{2n}\,\cos\alpha_{2nzx}\,I_{n},
\qquad n>0,
\end{equation}
\begin{equation}
2\tilde{\gamma}^{0}_{1}
=
\Delta g_{zx}^{0}\,I_{0},
\end{equation}
where
\begin{equation}
I_{n}=
\int_{0}^{2\pi}d\phi_{\mathrm{Tilt}}\,
\frac{\cos(2n\phi_{\mathrm{Tilt}})}{
\sqrt{
g_{x}^{2}\cos^{2}\phi_{\mathrm{Tilt}}
+
g_{y}^{2}\sin^{2}\phi_{\mathrm{Tilt}}}}
\left(
1-\frac{g_{x}g_{y}}{
g_{x}^{2}\cos^{2}\phi_{\mathrm{Tilt}}
+
g_{y}^{2}\sin^{2}\phi_{\mathrm{Tilt}}}
\right).
\end{equation}

Assuming $g_{x}>g_{y}$,
\begin{equation}
I_{n}=\frac{J_{n}(k)-\frac{g_{y}}{g_{x}}H_{n}(k)}{g_{x}},
\qquad
k=1-\frac{g_{y}^{2}}{g_{x}^{2}},
\quad
0<k<1.
\end{equation}

These integrals can be represented in terms of generalized hypergeometric functions ${}_2F_{1}$:
\begin{align}
J_{m}(k)
&=
2\pi(-1)^{m}\!\left(\frac{k}{4}\right)^{m}
\frac{\Gamma\!\left(m+\frac{1}{2}\right)}{\sqrt{\pi}\,\Gamma(m+1)}
{}_{2}F_{1}\!\left(m+\frac{1}{2},\,m+\frac{1}{2};\,2m+1;\,k\right),
\\
H_{m}(k)
&=
4\pi(-1)^{m}\!\left(\frac{k}{4}\right)^{m}
\frac{\Gamma\!\left(m+\frac{3}{2}\right)}{\sqrt{\pi}\,\Gamma(m+1)}
{}_{2}F_{1}\!\left(m+\frac{3}{2},\,m+\frac{1}{2};\,2m+1;\,k\right).
\end{align}

For $m=0$,
\begin{equation}
J_{0}(k)=4K(k),
\qquad
H_{0}(k)=\frac{4\,E(k)}{1-k},
\end{equation}
where $K$ and $E$ are complete elliptic integrals of the first and second kind.

\subsection{Numerical simulation details}

\begin{figure}[h!]
  \centering
  \includegraphics[width=0.6\textwidth]{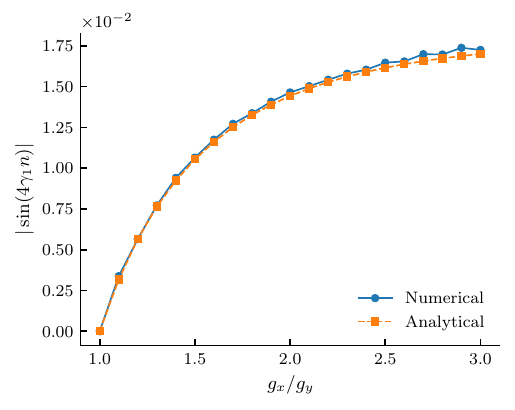}
  \caption{Numerical (blue) vs analytical (orange) expectation value $|\langle Y \rangle|$ as a function of $g_x/g_y$ for fixed $g_y=0.5$. The evolution is simulated for an in-plane magnetic field $B_x=60~\mathrm{mT}$ over a sweep duration $T_{\mathrm{phys}}=50~\mathrm{ns}$ and repeated for $n_{\mathrm{cycles}}=60$ cycles. The $g$-tensor modulation amplitude is set by $\delta=5\times10^{-4}$ (with an additional constant offset $\delta_1=3\times10^{-5}$), with the order of magnitude in accordance with having $\sim 10$ TLFs. The analytical curve is obtained from Eq.~\eqref{eq:gamma}.}
  \label{fig:supp_expectation}
\end{figure}

All numerical data are obtained by direct integration of the Schr\"odinger equation under the full time-dependent Hamiltonian $H(t)=H[\phi_{\mathrm{Tilt}}(t)]$ defined above in Eq.~\eqref{hamilt_prot}, with $\mathbf g(\phi_{\mathrm{Tilt}})=\hat g_{\parallel}(\phi_{\mathrm{Tilt}})+\Delta\mathbf g(\phi_{\mathrm{Tilt}})$ and $\hat g_{\parallel}(\phi_{\mathrm{Tilt}})=\hat R(\phi_{\mathrm{Tilt}})\,\mathrm{diag}(g_x,g_y,g_{\perp})\,\hat R^{T}(\phi_{\mathrm{Tilt}})$ (rotation about $z$). The perturbation is implemented as a tilt-dependent matrix with a single scalar modulation
\begin{equation}
\Delta\mathbf g(\phi)=
\begin{pmatrix}
0 & d(\phi)/2 & d(\phi)\\
d(\phi)/2 & 0 & 0\\
d(\phi) & 0 & 0
\end{pmatrix},
\qquad
d(\phi)=\delta\cos(n\phi+\alpha)+\delta_2\cos(m\phi+\alpha_1)+\delta_1,
\end{equation}
with $(\delta,n,\alpha)=(5\times 10^{-4},\,2,\,0.3)$, $(\delta_2,m,\alpha_1)=(10^{-7},\,4,\,0.4)$, and $\delta_1=3\times 10^{-5}$. The tilt sweep is parametrized by a smooth cubic ramp $u(t)=3s^2-2s^3$ with $s=t/T$, namely
\begin{equation}
\phi_F(t)=\phi_0+2\pi u(t),\qquad
\phi_B(t)=\phi_0+2\pi\bigl[1-u(t)\bigr],
\end{equation}
so that the backward sweep is the time-reversed version of the forward one. For the 2D $(g_x,g_y)$ sweep in the main-text readout maps we start the cycle at a small finite angle, $\phi_0=0.2~\mathrm{rad}$, to improve numerical stability for the smallest gaps (small $g_x$), whereas for the 1D numerical-analytical comparison in Fig.~\ref{fig:supp_expectation} we use $\phi_0=0$.

For each parameter point $(g_x,g_y)$ we define the local eigenbasis at $\phi_{\mathrm{Tilt}}=\phi_0$ from the unperturbed Hamiltonian $H_0[\phi_0]$: if $|-\rangle_{\phi_0}$ and $|+\rangle_{\phi_0}$ are the instantaneous ground and excited states, we set the local Pauli operators $X=|+\rangle\langle-|+|-\rangle\langle+|$ and $Y=i|+\rangle\langle-|-i|-\rangle\langle+|$ in this basis. The initial state is prepared as $|\psi_0\rangle=(|-\rangle_{\phi_0}+|+\rangle_{\phi_0})/\sqrt{2}$ and the full tilt-echo block (forward sweep $\rightarrow X \rightarrow$ backward sweep $\rightarrow X$) is repeated $n_{\mathrm{cycles}}$ times; the reported plotted signal is $|\mathrm{Re}\langle Y\rangle|$ of the final state.

In the 2D parameter sweep shown in the main text we use $g_x,g_y\in[0.03,0.3]$ with step $0.01$, $n_{\mathrm{cycles}}=10$, and a fixed sweep duration $T_{\mathrm{phys}}=1~\mu\mathrm{s}$ and $B_{\mathrm{phys}}=200~\mathrm{mT}$. For Fig.~\ref{fig:supp_expectation} we instead fix $g_y=0.5$, $B_{\mathrm{phys}}=60~\mathrm{mT}$ and $T_{\mathrm{phys}}=0.5\times 10^{-7}~\mathrm{s}$, vary $g_x$ (plotted as $g_x/g_y$), and use $n_{\mathrm{cycles}}=60$. The analytical curve in Fig.~\ref{fig:supp_expectation} is computed from Eq.~\eqref{eq:gamma} using the same $d(\phi)$ and plotted as $|\sin\!\bigl(4 n_{\mathrm{cycles}}\tilde{\gamma}_{1}\bigr)|$.

\section{Filter-function description of the tilt-echo protocol}
\label{sec:filter_function}

\subsection{Temporal fluctuations during repeated tilt-echo measurements.}

The directly measured observable of the protocol is the final expectation value
\begin{equation}
\langle Y \rangle = \sin \Phi,
\label{eq:sigy_obs}
\end{equation}
where \(\Phi\) is the total relative phase accumulated during the echoed tilt cycle. We consider a repeated-measurement experiment in which the same tilt-echo sequence is executed many times, while a nearby fluctuator switches stochastically between its two metastable states from run to run and, possibly, during a run. The experimentally relevant signal is therefore the average of Eq.~\eqref{eq:sigy_obs} over realizations of the fluctuator dynamics.

For \(\vec{B}=(B_x,0,0)\), the effective Hamiltonian takes the form
\begin{equation}
H_{\mathrm{eff}}(t)=\frac{\mu_B B_x}{2}
\left[
g_{xx}(t)\sigma_x+g_{yx}(t)\sigma_y+g_{zx}(t)\sigma_z
\right].
\label{eq:Heff_Bx}
\end{equation}
A single fluctuator modifies all three coefficients simultaneously. Since the tilt angle \(\phi \equiv \phi_{\mathrm{Tilt}}\) is swept during the protocol, each correction generally contains angular harmonics,
\begin{equation}
\Delta g_{ij}(\phi,t)=\sum_{m\ge 0}\Delta g_{ij}^{(m)}(t)\cos\!\bigl(m\phi+\alpha_{ij}^{(m)}\bigr),
\qquad ij\in\{xx,yx,zx\}.
\label{eq:gij_harmonics}
\end{equation}
We separate each harmonic into a static contribution and a fluctuating part,
\begin{equation}
\Delta g_{ij}^{(m)}(t)=\bar g_{ij}^{(m)}+u_{ij}^{(m)}(t),
\label{eq:gij_split}
\end{equation}
where \(\bar g_{ij}^{(m)}\) contributes to the static accumulated phase, while \(u_{ij}^{(m)}(t)\) generates temporal phase noise. We model the fluctuating part by a symmetric random telegraph process \cite{PaladinoRMP2014}
\begin{equation}
\xi(t)=\pm 1,
\end{equation}
switching with rate \(\Gamma\), and write
\begin{equation}
u_{ij}^{(m)}(t)=v_{ij}^{(m)}\,\xi(t),
\label{eq:RTNcoupling}
\end{equation}
where \(v_{ij}^{(m)}\) is the coupling amplitude of the fluctuator to harmonic \(m\) of channel \(ij\). The corresponding power spectral density is
\begin{equation}
S_{\xi}(\omega)=\frac{4\Gamma}{\omega^2+4\Gamma^2}.
\label{eq:RTNspec_base}
\end{equation}
Because the same fluctuator contributes to all three channels, the stochastic corrections to the geometric and dynamical parts of the accumulated phase are generally correlated.

The total phase accumulated during one echoed cycle can be written as, assuming perfect adiabaticity,
\begin{equation}
\Phi = 4\bar{\gamma}_1 + \delta\Phi,
\label{eq:Phi_total}
\end{equation}
where \(4\bar{\gamma}_1\) contains the static contribution to the accumulated phase, including the time-independent part of the fluctuator-induced correction, and \(\delta\Phi\) is the stochastic correction generated by the temporal fluctuations. To linear order, only the fluctuating part \(u_{ij}^{(m)}(t)\) contributes to \(\delta\Phi\), and we write
\begin{equation}
\delta\Phi
=
\delta\Phi_{zx}
+\delta\Phi_{xx}
+\delta\Phi_{yx},
\label{eq:dPhi_split}
\end{equation}
with
\begin{equation}
\delta\Phi_{ij}
=
\sum_m v_{ij}^{(m)}
\int_0^T dt\, y_{ij}^{(m)}(t)\,\xi(t),
\qquad ij\in\{zx,xx,yx\},
\label{eq:dPhi_channels}
\end{equation}
where \(T\) is the duration of one echoed cycle and \(y_{ij}^{(m)}(t)\) is the sensitivity kernel of that channel. The \(zx\) channel is geometric, whereas the \(xx\) and \(yx\) channels originate from imperfect cancellation of the dynamical phase.

The averaged measured signal is
\begin{equation}
\overline{\langle Y\rangle}
=
\overline{\sin\!\bigl(4\bar{\gamma}_1+\delta\Phi\bigr)},
\label{eq:avg_signal_exact}
\end{equation}
where the overline denotes averaging over repeated experimental runs. For telegraph noise, the exact average is generally non-Gaussian. In the regime considered here, however, the typical stochastic phase accumulated over the protocol remains small, \(\sqrt{\overline{\delta\Phi_n^2}} \ll 1\), such that higher cumulants are negligible. One may then approximate the averaged readout by
\begin{equation}
\overline{\langle Y\rangle}
\approx
\sin(4\bar{\gamma}_1)\,e^{-\chi_1/2}
\label{eq:avg_signal_gauss}
\end{equation}
for one echoed cycle, and more generally
\begin{equation}
\overline{\langle Y\rangle}
\approx
\sin(4n\bar{\gamma}_1)\,e^{-\chi_n/2}
\label{eq:avg_signal_gauss_n}
\end{equation}
after \(n\) repeated echoed cycles.

It is then convenient to introduce a filter-function representation for the total stochastic phase. For the zero angular harmonic and \(\vec{B}=(B_x,0,0)\), the one-cycle kernels are
\begin{equation}
y_{zx}^{(0)}(t)=4|\dot{\phi}(t)|\,\mathcal{K}(\phi(t)),
\label{eq:yzx0}
\end{equation}
\begin{equation}
y_{xx}^{(0)}(t)=
\frac{\mu_B B_x}{\hbar}\,
s(t)\,
\frac{g_{\parallel,xx}(\phi(t))}{D(\phi(t))},
\qquad
y_{yx}^{(0)}(t)=
\frac{\mu_B B_x}{\hbar}\,
s(t)\,
\frac{g_{\parallel,yx}(\phi(t))}{D(\phi(t))},
\label{eq:yxxyx0}
\end{equation}
where
\begin{equation}
g_{\parallel,xx}(\phi)=g_x\cos^2\phi+g_y\sin^2\phi,
\qquad
g_{\parallel,yx}(\phi)=(g_x-g_y)\sin\phi\cos\phi,
\end{equation}
\begin{equation}
D(\phi)=\sqrt{g_x^2\cos^2\phi+g_y^2\sin^2\phi},
\qquad
\mathcal{K}(\phi)=
\frac{1}{2D(\phi)}
\left(
1-\frac{g_xg_y}{D^2(\phi)}
\right),
\end{equation}
and
\begin{equation}
s(t)=
\begin{cases}
+1, & 0\le t\le T/2,\\
-1, & T/2\le t\le T.
\end{cases}
\end{equation}

For a linear forward sweep and palindromic return,
\begin{equation}
\phi(t)=
\begin{cases}
\Omega t, & 0\le t\le T/2,\\[2mm]
\Omega (T-t), & T/2\le t\le T,
\end{cases}
\qquad
\Omega=\frac{4\pi}{T}.
\label{eq:linear_sweep}
\end{equation}
Higher angular harmonics are incorporated by multiplying the zero-harmonic kernels by the corresponding modulation factor,
\begin{equation}
y_{ij}^{(m)}(t)=y_{ij}^{(0)}(t)\cos\!\bigl(m\phi(t)+\alpha_{ij}^{(m)}\bigr),
\qquad ij\in\{zx,xx,yx\}.
\label{eq:ym}
\end{equation}
Since the same telegraph process drives all channels and harmonics simultaneously, the total kernel is
\begin{equation}
y_{\mathrm{tot}}(t)=\sum_{ij,m} v_{ij}^{(m)}\,y_{ij}^{(m)}(t),
\qquad ij\in\{zx,xx,yx\},
\label{eq:ytot}
\end{equation}
and the corresponding filter function for one echoed cycle is
\begin{equation}
Y_{\mathrm{tot}}(\omega)=\int_0^T dt\, y_{\mathrm{tot}}(t)e^{i\omega t},
\qquad
F_{\mathrm{tot}}(\omega)=|Y_{\mathrm{tot}}(\omega)|^2.
\label{eq:Ftot}
\end{equation}

Within the same second-cumulant approximation,
\begin{equation}
\chi_1=
\int_0^\infty \frac{d\omega}{\pi}\,
S_{\xi}(\omega)\,F_{\mathrm{tot}}(\omega),
\label{eq:chi1}
\end{equation}
while for \(n\) repeated echoed cycles one analogously has
\begin{equation}
\chi_n=
\int_0^\infty \frac{d\omega}{\pi}\,
S_{\xi}(\omega)\,F_{\mathrm{tot},n}(\omega).
\label{eq:chin}
\end{equation}
Defining
\begin{equation}
Y_{ij}^{(m)}(\omega)=
\int_0^T dt\, y_{ij}^{(m)}(t)e^{i\omega t},
\label{eq:Yijm}
\end{equation}
one may write
\begin{equation}
Y_{\mathrm{tot}}(\omega)=\sum_{ij,m} v_{ij}^{(m)}\,Y_{ij}^{(m)}(\omega),
\qquad
F_{\mathrm{tot}}(\omega)=\left|Y_{\mathrm{tot}}(\omega)\right|^2.
\label{eq:Ytot_decomp}
\end{equation}
For an approximately linear sweep, the main spectral weight of \(Y_{ij}^{(m)}\) is shifted toward frequencies of order \(m(4\pi/T)\).

Finally, for \(n\) repeated echoed cycles, where each cycle has duration \(T\),
\begin{equation}
Y_{\mathrm{tot},n}(\omega)=Y_{\mathrm{tot}}(\omega)\,
e^{i\omega(n-1)T/2}
\frac{\sin(n\omega T/2)}{\sin(\omega T/2)},
\end{equation}
so that
\begin{equation}
F_{\mathrm{tot},n}(\omega)=
F_{\mathrm{tot}}(\omega)
\left[
\frac{\sin(n\omega T/2)}{\sin(\omega T/2)}
\right]^2.
\label{eq:Fn}
\end{equation}
This result follows from the fact that \(n\) identical cycles correspond to a sum of \(n\) time-shifted copies of the same kernel, whose Fourier transform gives the usual geometric-series prefactor. Substituting Eq.~\eqref{eq:Fn} into Eq.~\eqref{eq:chin} gives the dephasing envelope of the repeated protocol.

\begin{figure}[t]
    \centering
    \includegraphics[width=\linewidth]{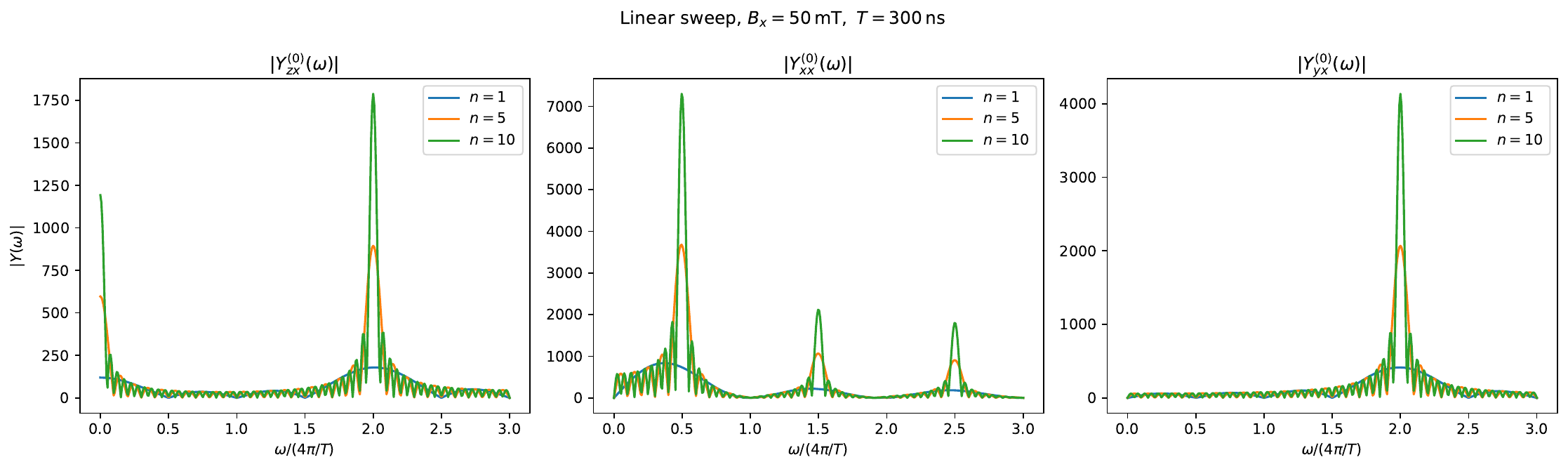}
    \caption{Zero-harmonic Fourier kernels \( |Y^{(0)}_{ij}(\omega)| \) for \(ij\in\{zx,xx,yx\}\) under a linear tilt sweep, for \(B_x=50\,\mathrm{mT}\), \(T=300\,\mathrm{ns}\), and \(n=1,5,10\) echoed cycles.}
    \label{fig:filter_kernels_threepanels}
\end{figure}

\subsection{Robustness of the tilt-echo protocol to miscalibrations.}
\label{sec:miscal}
Here we show that the tilt-echo protocol derived in Sec.~\ref{app:berry} is robust against weak time-dependent noise in the $g$ tensor, provided adiabaticity is preserved and the measured signal $\langle Y\rangle$ is averaged over many realizations of the same protocol. More precisely, if the gates and initial state are calibrated for a Hamiltonian with a slightly different $g$ tensor, the resulting miscalibration enters $\langle Y\rangle$ only at higher order.
At any time \(t\), we define the operator \(R(t)\) that relates the instantaneous eigenstates \(\{|\pm\rangle_{\mathrm{inst}}(t)\}\) of the actual Hamiltonian to the calibration basis:
\begin{equation}
|\pm\rangle_{\mathrm{inst}}(t)=R(t)|\pm\rangle_c.
\end{equation}
We assume that the mismatch between the instantaneous and calibration bases is small and denote its overall scale by \(\eta\). Using the same gauge convention for the local eigenstates as above \eqref{gauge}, the infinitesimal rotation in the local calibration frame is
\begin{equation}
R(t)=1-\frac{i}{2}\bigl(d\tilde{\theta}(t)\,Y-\sin\tilde{\theta}(t)\,d\tilde{\phi}(t)\,X+\cos\tilde{\theta}(t)\,d\tilde{\phi}(t)\,Z\bigr)+O(\eta^2).
\end{equation}
For the reference tensor \(\hat g_{\parallel}\) used in the main text, one has \(g_{zx}=0\) by construction. The longitudinal component \(\eta_{fz}=\cos\tilde{\theta}\,d\tilde{\phi}\) is therefore of order \(O(\eta^2)\) and can be neglected. To leading order, the basis mismatch is purely transverse, so we write
\begin{equation}
R(t)\approx 1+i\eta K(t),
\qquad
K(t)\in \mathrm{span}\{X,Y\}.
\end{equation}

We account for the basis mismatch at three points: \(R_0\) at the start (\(t=0\)), \(R_1\) at the midpoint (\(t=T/2\)), and \(R_2\) at the end (\(t=T\)). The forward and backward physical propagators are
\begin{equation}
U_f=R_1 e^{i\alpha_f Z} R_0^\dagger,
\qquad
U_b=R_2 e^{-i\alpha_b Z} R_1^\dagger,
\qquad
\alpha_{f,b}=\tilde{\gamma}_1 \pm \varepsilon_{f,b}.
\end{equation}
The full echo operator is \(S=XU_bXU_f\), constructed using calibrated \(X\) gates. Using the identity \(X e^{-i\alpha_b Z} X=e^{i\alpha_b Z}\) and \(\bar{A}=XAX\), we have
\begin{equation}
S=X (R_2 e^{-i\alpha_b Z} R_1^\dagger) X (R_1 e^{i\alpha_f Z} R_0^\dagger)
= \bar{R}_2 e^{i\alpha_b Z} \bar{R}_1^\dagger R_1 e^{i\alpha_f Z} R_0^\dagger.
\end{equation}
Defining \(R_k=1+i\eta K_k\) and expanding to first order in \(\eta\), we obtain
\begin{equation}
S \approx (1+i\eta \bar{K}_2)e^{i\alpha_b Z}(1+i\eta (K_1-\bar{K}_1))e^{i\alpha_f Z}(1-i\eta K_0)
= S_0(1+i\eta M)+O(\eta^2),
\end{equation}
where
\begin{equation}
S_0=e^{i(\alpha_f+\alpha_b)Z}.
\end{equation}
The total deviation operator is
\begin{equation}
M=e^{-i\alpha_f Z}(K_1-\bar{K}_1)e^{i\alpha_f Z}+S_0^\dagger \bar{K}_2 S_0-K_0.
\end{equation}
Since \(K_1\), \(\bar{K}_1\), \(\bar{K}_2\), and \(K_0\) are all in \(\mathrm{span}\{X,Y\}\), and this subspace is invariant under \(Z\)-rotations, \(M\) remains purely transverse.

We now consider the measured signal
\begin{equation}
\langle Y \rangle = \mathrm{Tr}(Y S \rho_0 S^\dagger),
\end{equation}
where \(\rho_0\), \(X\), and \(Y\) are defined in the calibration basis \(\{|\pm\rangle_c\}\). To first order in \(\eta\),
\begin{equation}
\langle Y \rangle \approx \mathrm{Tr}(Y S_0 \rho_0 S_0^\dagger) + i\eta\,\mathrm{Tr}(Y S_0 [M,\rho_0] S_0^\dagger) + O(\eta^2).
\end{equation}
Because \(M\in \mathrm{span}\{X,Y\}\) and \(\rho_0\approx (1+X)/2\), the commutator \([M,\rho_0]\) is proportional to \(Z\). Since \(S_0\) is a \(Z\)-rotation, \(S_0 Z S_0^\dagger=Z\), and the linear term vanishes because \(\mathrm{Tr}(Y Z)=0\).

Thus the signal is robust against this time-dependent basis mismatch:
\begin{equation}
\langle Y \rangle = \sin\left(4\tilde{\gamma}_1 + 2(\varepsilon_f - \varepsilon_b)\right) + O(\eta^2).
\end{equation}

\section{resonator coupling}
For readout, the spin can be shuttled either to a double quantum dot for Pauli-spin-blockade readout or to a double quantum dot operated in the flopping-mode regime and coupled to a microwave cavity. In either case, a recovery gate is applied to correct for spin evolution during transport and the inherent misalignment between the initial and final quantization axes. For the cavity-based implementation, this Supplemental Material provides an estimate of the integration time required to reach $\mathrm{SNR}\sim 1$.

The Hamiltonian for the spin-photon coupling takes the form:

\begin{align}
    H_{sp}=\omega_r a^\dagger a + \frac{\omega_q}{2}\tau_z +\eta(a^\dagger \tau_{-}+a \tau_{+}).
\end{align}
Where $\omega_r$ and $\omega_q$ are the resonator and qubit frequencies, $a$ and $\tau_z$ are the annihilation operator of the cavity mode and the Pauli operator for the spin qubit, and $\eta$ denotes the transverse spin–photon coupling strength, typically on the order of $10~\mathrm{MHz}$ \cite{stefanosselfcitation}.  Here we already neglected fast-oscillating terms such as $a^\dagger \tau_+$, as well as any longitudinal coupling term $\eta_{\parallel}\tau_z (a^{\dagger}+a)$, under the assumptions $|\omega_q - \omega_r| \ll \omega_q + \omega_r$ and $|\eta_{\parallel}/\omega_r| \ll \eta/|\omega_q - \omega_r|$. In the dispersive regime defined by $\Delta = \omega_q - \omega_r$ and $|\eta|\sqrt{n_a} \ll |\Delta|$, where $n_a$ is the average photon number, a Schrieffer–Wolff transformation yields:
\cite{didierfast,gambetta2008quantum}
\begin{align}
    H_{sp} \approx (\omega_r+\chi\tau_z)a^\dagger a+\tfrac{1}{2}(\omega_q+\chi)\tau_z,
\end{align}

\begin{align}
    \chi=\frac{\eta^2}{\Delta},\quad
n_{\rm crit}=\frac{\Delta^2}{4\eta^2}.
\end{align}
We now consider the cavity dynamics under continuous driving, using standard input–output formalism. $\gamma$ denotes the cavity decay rate through the measurement port. Assuming the drive is resonant with the bare cavity mode ($\omega_d = \omega_r$), the Langevin equation for $a$ in the interaction picture reads \cite{didierfast}:
\[
\dot{a} = -i\chi \tau_z a - \frac{\gamma}{2} a - \sqrt{\gamma}\, a_{\mathrm{in}},
\]
with the output field given by
\[
a_{\mathrm{out}} = a_{\mathrm{in}} + \sqrt{\gamma}\, a.
\]
With a drive of amplitude $\epsilon\equiv|\epsilon|$, we write the input field as
$\hat a_{\mathrm{in}}=\alpha_{\mathrm{in}}+\hat d_{\mathrm{in}}$, with average
$\alpha_{\mathrm{in}}=\langle \hat a_{\mathrm{in}}\rangle=-\,\epsilon/\sqrt{\gamma}$ and fluctuations
$\hat d_{\mathrm{in}}=\hat a_{\mathrm{in}}-\alpha_{\mathrm{in}}$.
To read out the qubit, we perform homodyne detection of the output field and integrate the chosen quadrature over some measurement time $\tau_m$:
\[
\hat M(\tau_m)
=\sqrt{\gamma}\int_{0}^{\tau_m}\!dt\,\big[i\,\hat a_{\mathrm{out}}(t)+\mathrm{h.c.}\big].
\]
The SNR is defined as
$
    \mathrm{SNR}^2 \equiv \frac{\left|\langle\hat{M}(\tau_m)\rangle\right|^2}{\left\langle\hat{M}_{N }^2(\tau_m)\right\rangle},
$
where
$ \hat{M}_N=\hat{M}-\langle\hat{M}\rangle$. For $\tau_m\gamma\gg1$, $\langle \hat{a}_{out}\rangle$ reaches a steady state value and $\hat{M}(\tau_m)$ becomes:
\begin{align}
    \hat{M}(\tau_m)_{ss}= 2\epsilon \langle \tau_z \rangle \tau_m\sin \phi_{qb}, \ \ \ \phi_{qb} = 2\arctan\frac{2\chi}{\gamma}.
\end{align}
The SNR is maximized for the optimal detuning $2\chi=\gamma$ (which, for $\gamma \sim 1 \ \text{MHz}$, can be achieved at detunings $\sim \Delta$ of fractions of GHz):
\begin{align}
    \text{SNR}_{\tau_m  \gamma \gg 1} = 2\epsilon|\langle \tau_z \rangle|\sqrt{\frac{\tau_m}{\gamma}}.
\end{align}
\par A recovery gate is applied after shuttling such that we take $\langle \tau_z\rangle \approx \langle \sigma_y\rangle_p$ at the end of the pulse sequence. For the typical $\delta\hat g$ variations obtained in simulation, $|\langle\tau_z\rangle |\sim 10^{-1}-10^{-2}$ is reached within $\sim 1-10 \mu$s. Since the steady-state intra-cavity photon number is $n_{\mathrm{ss}}=4 |\epsilon|^2/\gamma^2$, choosing a drive such that $n\sim n_{\mathrm{crit}}/10$ yields an SNR $\sim 1$ in $\sim 100 \text{ns} - 1\mu \text{s}$ \footnote{We neglect Purcell decay here; for our parameters the rate is $\Gamma_{\mathrm{P}}=\eta^2\gamma/\Delta^2\sim\mathrm{kHz}$ and the relaxation time will be likely dominated by phonons.}.

\end{document}